\documentclass[11pt]{article}
\usepackage[utf8]{inputenc}
\usepackage{arydshln}
\usepackage[]{bm}
\usepackage[]{amsmath}
\usepackage{rotating}
\usepackage[]{verbatim}
\usepackage{amsthm}
\usepackage{afterpage}
\usepackage{enumerate}
\usepackage[inline,shortlabels]{enumitem}
\usepackage[flushleft]{threeparttable}
\usepackage{mathtools}

\DeclareMathOperator*{\argmin}{arg\,min}
\usepackage[titletoc]{appendix}
\usepackage{epstopdf} 
\usepackage{graphicx}
\usepackage{lscape}
\usepackage{amssymb}
\usepackage[capposition=top]{floatrow}
\usepackage[left=0.9in, right=0.9in, top=1in, bottom=1in]{geometry}
\usepackage[normalem]{ulem}

\usepackage{tabularx,booktabs}
\usepackage{longtable}
\usepackage{booktabs}
\usepackage{graphicx}
\usepackage{multirow}
\usepackage{adjustbox}
\usepackage{array}
\newcolumntype{H}{>{\setbox0=\hbox\bgroup}c<{\egroup}@{}}
\usepackage{tikz}
\usetikzlibrary{shapes.geometric, arrows,chains}
\usetikzlibrary{calc, shapes,positioning}
\usepackage{subcaption}
\usepackage{authblk}

\usepackage{setspace}
\usepackage{xr}
\usepackage{chngcntr}
\usepackage{apptools}
\AtAppendix{\counterwithin{lemma}{section}}
\AtAppendix{\counterwithin{assumption}{section}}
\usepackage{longtable}
\usepackage{algpseudocode}
\usepackage[ruled]{algorithm2e}
\allowdisplaybreaks
\algblock{Input}{EndInput}
\algnotext{EndInput}
\algblock{Output}{EndOutput}
\algnotext{EndOutput}

\DeclareMathOperator*{\plim}{plim}
\usepackage{indentfirst} 
\usepackage{titlesec}
\titlelabel{\thetitle.\hspace{.5em}} 
\usepackage[colorlinks=true, linkcolor=blue, citecolor=blue, urlcolor=blue]{hyperref}
\makeatletter
\renewcommand{\fnum@figure}{Fig. \thefigure}
\makeatother
\usepackage[labelfont=bf]{caption}
\ExplSyntaxOn
\NewDocumentEnvironment{sequation}{O{\small}b}
 {
 \yufip_sequation:nnn {equation}{#1}{#2}
 }{}
\NewDocumentEnvironment{sequation*}{O{\small}b}
 {
 \yufip_sequation:nnn {equation*}{#1}{#2}
 }{}
\cs_new_protected:Nn \yufip_sequation:nnn
 {
 \begin{#1}
 \mbox{#2$\displaystyle#3$}
 \end{#1}
 }
\ExplSyntaxOff

\tikzset{
  startstop/.style={
    rectangle, 
    rounded corners,
    minimum width=3cm, 
    minimum height=1cm,
    align=center, 
    draw=black, 
    fill=red!30
    },
  process/.style={
    rectangle, 
    minimum width=3cm, 
    minimum height=1cm, 
    align=center, 
    draw=black, 
    fill=blue!30
    },
  decision/.style={
    rectangle, 
    minimum width=3cm, 
    minimum height=1cm, align=center, 
    draw=black, 
    fill=green!30
    },
  arrow/.style={thick,->,>=stealth},
  dec/.style={
    ellipse, 
    align=center, 
    draw=black, 
    fill=green!30
    },
}

\usepackage[backend=biber,style=apa,natbib=true]{biblatex}
\begin{document}
\newgeometry{left=0.9in, right=0.9in, top=0.7in, bottom=1in}
\setlength{\abovedisplayskip}{5pt}
\setlength{\belowdisplayskip}{5pt}
\begin{spacing}{1.4}
\title{The Effect of Capital Share on Income Inequality: \\ Identifying the Time Patterns$^*$}

\author[1]{Oğuzhan Akgün}
\affil[1]{LEDi, University of Burgundy, France}

\author[2]{Ezgi Özsöğüt}
\affil[2]{Paris School of Economics, France}
\date{\today}
\maketitle
\vspace{-1cm}

\begin{abstract}
This study explores the link between the capital share and income inequality over the past four decades across 56 countries.
Calculating the capital share from national accounts alongside top income share data from the World Inequality Database, which is based on the Distributional National Accounts methodology,  we ensure the consistency in the theory and measurement. 
Employing a structural econometric approach, we account for heterogeneous and time-varying transmission coefficients from the capital share to personal income  inequality.
Our findings reveal that a one percentage point (pp) increase in the capital share raises the income share of the top 5\% by 0.17 pp on average.
Advanced economies show a stable transmission coefficient with rising capital and labor income inequality, while emerging economies experience an increasing transmission coefficient alongside growing capital income inequality.
In contrast, a third group exhibits a declining transmission coefficient and rising labor income inequality.
Overall, changes in the capital share account for approximately 50\% of the rise in income inequality, underscoring its pivotal role over the last four decades.
\vspace{0.3cm} \newline

\end{abstract}
\vspace{-0.5cm}

\rule{1.5in}{0.01in}

{\small
$^*$This paper developed as a chapter of the Ph.D. project of Ezgi Özsöğüt. We thank Ezgi Özsöğüt’s Ph.D. advisors, Florin Bilbiie and Romain Rancière, for their feedback on earlier versions of this paper, as well as jury members Luca Fornaro, Tobias Broer, and Mirko Wiederholt for their comments. Special thanks to Caroline Betts for her careful review and valuable suggestions as the main reviewer. We also thank the seminar participants at Galatasaray University Economic Research Center (October, 2024). The usual disclaimer applies.

Email:
oguzhan.akgun@u-bourgogne.fr (Corresponding author, O.\ Akgün)
}
\end{spacing}

\setcounter{page}{0}\thispagestyle{empty}\pagebreak 
\restoregeometry

\section{Introduction} \label{sec:introduction}

In a dual-class society, where the population is divided into wage-earning workers and wealthy capitalists who own capital, the link between personal income distribution and the capital-labor split in national income is straightforward.
However, such societies arguably no longer exist.
Although the capital remains concentrated in the hands of high income earners, in modern capitalist economies, different income groups earn both capital and labor incomes \citep{milanovic2017}.
This weakens the assumption of an obvious positive and strong relationship between the capital share and income inequality prompting us to ask some previously unremarkable questions: how does the capital share impact income inequality in different countries, and how does this relationship evolve over time?

Once the theoretical relationship linking the capital share to income inequality is established \citep[e.g.,][]{atkinson-bourgignon2000, atkinson2009, milanovic2017}, the most direct way to quantify this link is by applying an inequality decomposition using microdata sources \citep{lerman1985income, milanovic2002decomposing, amarante2016income, nolan2021intergenerational}.
These sources include surveys and what we refer to as fiscal data throughout the analysis, constructed using micro-level tax records or tax tabulations.
Relying on country-level data sources, some studies have successfully applied these methods \citep[e.g.,][]{giangregorio2024functional, aaberge2018, atkinson-lakner2021}, but extending the analysis to a cross-country framework is challenging due to the lack of high quality microdata in most countries.
Even when the data are available, there are usually significant methodological gaps both in fiscal data and surveys, complicating cross-country comparisons.

Another strand of the literature applies decompositions using harmonized surveys from the Luxembourg Income Study (LIS) \citep[e.g.,][]{garcia2013factor, francese2015functional}, but despite substantial efforts in standardizing definitions across countries, LIS data still faces important limitations in measuring income inequality accurately.
\citet{flores2021capital} shows that both surveys and fiscal data systematically underestimate capital income, and hence income inequality, due to \textit{measurement errors} and \textit{conceptual differences} between microdata and macrodata sources.
According to the author's calculations, surveys and fiscal data in the United States capture only 20\% and 30\% of capital incomes on average, respectively, while they capture up to 70\% and 80\% of labor incomes.
Such underestimation of capital incomes, even in a country with exceptionally high quality individual data, leads to biased estimates of income inequality, making decomposition-based analyses unreliable for exploring the effect of the capital share on income inequality.

The rest of the empirical literature, on the other hand, largely relies on reduced form approaches to estimate the effect of the capital share on income inequality \citep{daudey2007personal, francese2015functional, schlenker2015capital, bengtsson2018capital, erauskin2020labor}.
While these models offer flexibility, they often fail to capture the true structural relationship between the two variables.
This limits their capacity to explain the underlying mechanisms and complicates the interpretation of their findings \citep{flores2021capital}.
Furthermore, in most cases, the measurement inconsistency persists because the dependent variable, income inequality, is still calculated using microdata (i.e., surveys or fiscal data) and the right-hand side variable, the capital share, is derived from macrodata (i.e., national accounts).
\citet{bengtsson2018capital}, for instance, which is arguably the study most closely related to ours, relies on top income shares calculated based on \textit{fiscal income}.
Although these were the longest series covering a relatively large number of countries at the time of the study, they do not fully align with the income definitions used to calculate the capital shares from national accounts.
As the authors acknowledge, this inconsistency possibly bias their estimates but this limitation stems from the study’s focus on long-term relationships and the construction of historical series, which it successfully accomplishes.
On the empirical side, the study assumes a log-linear relationship between the capital share and inequality, which cannot fully capture the theoretical link between the two variables.
Dividing the sample into sub-periods provides further insights on parameter shifts over time but remains limited in addressing broader time-varying effects.

The main contribution of the present paper is building an empirical framework addressing these shortcomings of the previous empirical studies and closing the gap between the theoretical and the empirical literature.
Specifically, we directly estimate the structural relationship between the capital share and income inequality while ensuring the consistency in the measurement of both variables.
To measure inequality, we use the top income share series from the World Inequality Database (WID) calculated using the Distributional National Accounts (DINA) methodology proposed by \citet{dina2021}. 
When rigorously implemented, the DINA methodology ensures that \textit{all} types of income recorded in national accounts are properly accounted for and distributed to individuals, providing consistency in the income definitions used to construct capital shares and top income shares. 
The data set that we construct covers the last four decades of up to 56 emerging and advanced countries.
This effort of building a large and internally consistent data set of capital shares and income inequality represents a simple yet significant improvement in the measurement over the previous studies.

In terms of the econometric methodology, our contribution is the estimation of a structural equation, while accounting for the heterogeneous and time-varying nature of what we refer to as \textit{the transmission coefficient}, i.e., the effect of the capital share on income inequality.
To achieve this, we develop a two-step estimation procedure following the recent panel data literature such as \citet{bonhomme2019distributional,bonhomme2022discretizing}, \citet{su2019sieve} and \citet{chen2019estimating}.
In the first step, the countries in the sample are classified in groups of homogeneity via the Kmeans clustering algorithm, using the time averages of their top income and capital shares.
In the second step, these groups are used to estimate a model with time-varying coefficients via penalized least squares (PLS) following \citet{wood2000modelling,wood2006confidence,wood2017generalized}.
Each step of our methodology has a valid economic interpretation and reveals an important aspect of the relationship between the capital share and income inequality.
We also address the potential endogeneity of the capital share which may be present in an income inequality regression due to measurement errors.
For this purpose, we use an identification strategy based on an instrumental variable, the profit tax rate, which is found to be strongly correlated with the capital share.


The empirical findings can be summarized as follows.
\begin{enumerate*}[label=(\roman*)]
\item When the data from the 56 countries in the sample are pooled, our preferred estimate of the transmission coefficient equals 0.17.
This shows that one percentage point (pp) increase in the capital share increases the income share of the richest top 5\% of the population by 0.17 pp, on average over the period covered.
Incorporating the instrumental variable increases this estimate to 0.28.
\item For the full sample, the transmission coefficient rises between 1980 and 2000.
After 2000, it stabilizes with some fluctuations with a noteworthy increase by the end of the period.
These changes appear to be mainly driven by the changes in capital income inequality.
\item When different groups are considered, however, the time patterns show important heterogeneity.
In the group which is mainly made of advanced economies, with low inequality and capital shares, the transmission coefficient is found to be stable in the last four decades, while both capital and labor income inequality rise steadily until 2008.
In a group of mostly emerging economies with high inequality and high capital share, the transmission coefficient rises significantly along with capital income inequality.
In the third, more heterogeneous group of countries showing moderate levels of inequality and capital shares, the transmission coefficient declines, while labor income inequality rises throughout the period.
\item Finally, the Shapley decomposition of the model predictions reveals that the main driver of the changes in income inequality has been the changes in the capital share over the last four decades.
The percentage of the average increase in the income inequality due to the changes in capital shares is found to be around 50\% for most of the countries in the sample.
\end{enumerate*}

The rest of the paper is organized as follow.
In Section \ref{sec:theoreticalframework} the theoretical framework outlining the structural relationship between the capital share and income inequality is introduced.
Section \ref{sec:measurement} presents the variable choices and the data.
Section \ref{sec:empiricalmethodology} describes the estimation strategy.
In Section \ref{sec:results} we present our findings, and Section \ref{sec:conclusion} concludes.

\section{Theoretical Framework}\label{sec:theoreticalframework}

We start by outlining the theoretical link between personal income inequality and the capital share.
This link relies on an accounting identity which decomposes inequality measures, such as the total income Gini, as in \citet{milanovic2017}, or income shares, as in \citet{atkinson-lakner2021}, into factor shares and their distribution throughout the population.

Let $Y_{q,m}$ be the income of the top $q\%$ of the population obtained from factor $m \in \{K,L\}$ where $K$ and $L$ represent capital and labor, respectively.
We denote the special case of the total income from factor $m$ as $Y_{\circ,m} := Y_{100,m}$, and similarly we write $Y_{q,\circ} := \sum_{m \in \{K,L\}} Y_{q,m}$.
The total income in the economy will be shown simply as $Y$, that is $Y := Y_{100,\circ}$.
The income share of the top $q\%$, defined as $S^0_{q,\circ} = Y_{q,\circ}/Y$, can then be written as
\begin{equation*}
      S^0_{q,\circ} = S^0_{q,K} \times CS^0 + S^0_{q,L} \times LS^0
\end{equation*}
where $CS^0 = Y_{\circ,K}/Y$ and $LS^0 = Y_{\circ,L}/Y$ are the capital and labor shares in the economy, and $S^0_{q,m} = Y_{q,m}/Y_m$ is the share of the total income from factor $m$ received by the top $q\%$ of the population.
Throughout the paper, we use the superscript $^0$ to denote the true variables, in order to make the distinction between the theoretical quantities and their observed counterparts.
In this identity, $S^0_{q,\circ}$ captures the personal income inequality of the economy, and $S^0_{q,m}$ measures the inequality in the marginal distributions of factor incomes, that is, labor income inequality and capital income inequality.

Since the capital share and the labor share sum up to one, the top income share of quantile $q$ can be written as
\begin{equation} \label{eq:estimate}
    S^0_{q,\circ} = D^0_q \times CS^0 + S^0_{q,L}
\end{equation} 
where $D^0_q = S^0_{q,K} - S^0_{q,L}$.
This identity provides a useful description of the link between income inequality and the capital share.
Namely, it shows that the personal income inequality, as measured by $S^0_{q,\circ}$, can be explained by three factors:
\begin{enumerate*}[(i)]
    \item the distribution of the capital income among individuals, measured by $S^0_{q,K}$,
    \item the distribution of the labor income among individuals, measured by $S^0_{q,L}$,
    \item and the distribution of the income between labor and capital, that is, the functional income distribution of the economy, measured by $CS^0$.
\end{enumerate*}
We therefore see that the marginal effect of the capital share depends not only on a more or less concentrated distribution of capital income across individuals ranked in the top of the income distribution, but also by the difference between the concentration parameters of the capital and the labor income distributions. 

Equation \eqref{eq:estimate} also provides a potential explanation for why the relationship between the capital share and income inequality received little attention until recently.
Historically a high concentration of capital incomes in favor of the rich (high $S^0_{q,K}$) and a high concentration of labor incomes in favor of people with low overall income (low $S^0_{q,L}$) were both assumed as obvious facts, providing a sufficiently high $D^0_{q}$ indicating a positive and strong relationship between the capital share and inequality.
\citet{ranaldi2022capitalist} describe the extreme case as \textit{classical capitalism} in which income composition inequality is at its maximum, that is, the rich earn only capital income while the rest earns only labor income.
This corresponds to the scenario with $S^0_{q,K}=1$ and $S^0_{q,L}=0$ in Equation \eqref{eq:estimate}.
In this case $D^0_{q} = 1$ and changes in the functional distribution mirrors changes in personal income distribution.
The evidence provided by the authors show that this is not anymore the case. 
Although the capital income remains relatively concentrated in favor of the overall rich, in modern capitalist economies both labor and capital incomes are distributed across different income groups.
Hence, it is likely that $D^0_{q} \neq 1 $ with significant heterogeneity across countries.
\citet{aaberge2018} and \citet{atkinson-lakner2021} present similar results for Norway and the United States respectively.
Overall, the evidence suggests that the positive and strong relationship between the capital shares and top income shares is not as obvious as it once was, and changes in the capital share is no longer sufficient to explain changes in income distribution.

With the theoretical link between the capital share and inequality established, we now turn to quantifying this relationship.
The next section outlines our measurement approach and the data set.

\section{Data Set}\label{sec:measurement}

In line with DINA guidelines and following standard practice in literature \citep[see][among others]{bengtsson2018capital}, we measure total capital income $Y_{\circ,K}$ as the sum of net national capital income (net of consumption of fixed capital) and 30\% of mixed income (i.e., self employment income).
This definition includes all types of capital income recorded in national accounts which should be attributed to resident individuals.  
For instance, owner occupied housing income as well as all types of financial incomes, whether they are directly received in the form of dividends by households or retained in corporations in forms of undistributed profits, are included. 
Total income corresponds to net national income at factor cost, that is, sum of capital, labor and mixed incomes, net of consumption of fixed capital.
This definition also excludes net taxes on products, which is standard practice in literature measuring factor shares, as these are indirect taxes paid to the government before they could be attributed to any factor of production \citep{dina2021}.
The capital share used in this study is then obtained as the ratio of these two variables,
corresponding to the empirical counterpart of $Y_{\circ,K}/Y$ in the theoretical model.

In order to measure income inequality accurately, in a way that aligns with the income definitions used to calculate the capital share, we avoid relying on measures derived from raw microdata, a key limitation of previous studies.
\citet{flores2021capital} provides an extensive review of the literature before quantifying the measurement gaps in these studies, and identifies two key factors causing both fiscal data and surveys to underestimate capital incomes, hence income inequality.
First, total national income, as defined by the System of National Accounts (SNA), includes income flows beyond household incomes, such as undistributed corporate profits and imputed rents for owner-occupied housing.
These capital income flows, in reality, belong to individuals.
Thus, they are expected to contribute to overall inequality.
They are, however, often excluded from personal income definitions in surveys or fiscal data used to calculate inequality series while included in the calculation of capital shares from national accounts.
This is the first source of measurement concern labeled as \textit{conceptual differences}.
Second, even when focusing solely on household incomes, microdata often suffer from under-reporting (fiscal data) and sampling biases (surveys). Such omissions, affecting primarily capital incomes and those at the top of the distribution, result in \textit{systematic measurement errors}, which is the second source of mismeasurement present in previous studies.

To address these concerns, we use top income shares from WID, which are calculated using the DINA methodology \citep{dina2021}. Specifically, we extract Top 10\%, Top 5\%, and Top 1\% pre-tax income share series from WID, that is, $q \in \{ 10, 5, 1 \}$ in Equation \eqref{eq:estimate}. 
First, integrating various methods and data sources such as national accounts, surveys, tax records, and demographic structure, the DINA methodology distributes the entire national income to individuals.
In doing so, the calculated top income shares include all forms of income recorded in the national accounts and are not subject to the conceptual differences seen in traditional microdata.
Second, using additional data sources, the DINA methodology also applies corrections to account for measurement errors, which may be due to under-reporting or tax evasion, ensuring that income totals are correctly measured and distributed to individuals.
Therefore, when rigorously implemented, the DINA methodology ensures that \textit{all} types of income recorded in national accounts are properly accounted for in calculating inequality series, making them appropriately comparable with capital share series derived from national accounts. This represents a significant improvement in measurement compared to previous studies.

\begin{table}
  \centering
  \caption{Summary of the data set}
    \begin{tabular}{lll}
    \toprule
    Variable & Symbol & Source \\
    \midrule
    Capital share                   & $CS_{it}$                        & Own calculations using national accounts \\
    Top 10\%,  5\%, 1\% income shares    & $S_{it}$ & World Inequality Database \\
    Profit tax rate                 & $PTR_{it}$                       & World Development Indicators \\
    \bottomrule
    \end{tabular}%
  \label{tab:data_summary}%
\end{table}%

Note that, although the DINA methodology accounts for the entire national income in calculating income share series, non-personal components of income, such as net taxes on products, are distributed to individuals in proportion to their shares in factor-price national income.
In this way the overall inequality in total factor income and net national income is quantitatively equivalent.
Hence, using the capital share definition at factor cost (i.e., excluding net taxes on products) together with income shares calculated according to the DINA methodology remains fully consistent, ensuring that the accounting identity outlined in the previous section holds.

Finally, the profit tax rate, which is used as an instrument for the capital share, is obtained from World Development Indicators (WDI).
The resulting dataset, summarized in Table \ref{tab:data_summary}, covers an unbalanced panel of 56 countries from 1980 to 2020 with 1032 observations in total.
The profit tax rate is available only from 2005, thus covering only the last 16 years of the sample.

\begin{figure}[H]
    \centering
    \begin{subfigure}{0.9\textwidth}  
        \caption{Top income shares}  
        \includegraphics[width=\textwidth]{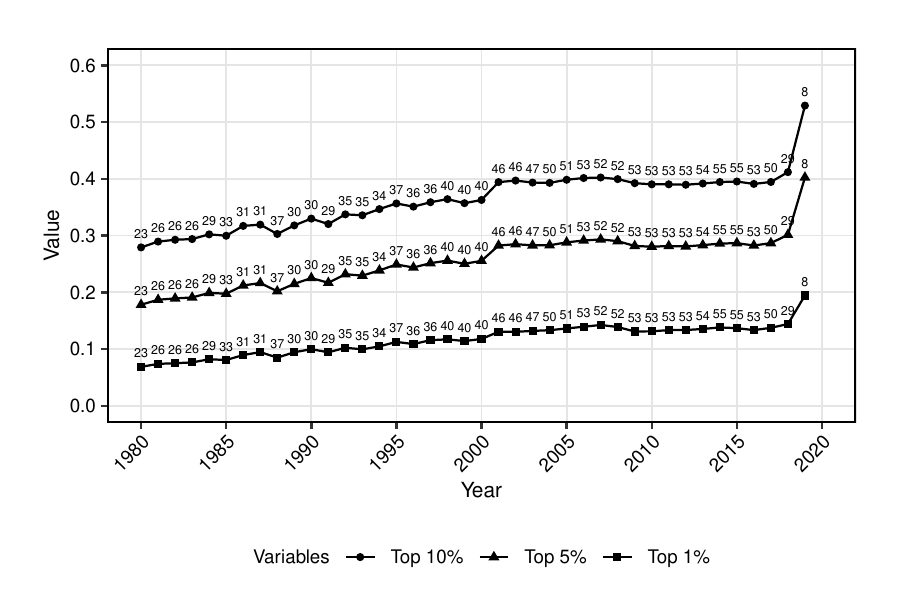}  
    \end{subfigure}
    \begin{subfigure}{0.9\textwidth}  
        \caption{Capital share and Profit tax rate}  
        \includegraphics[width=\textwidth]{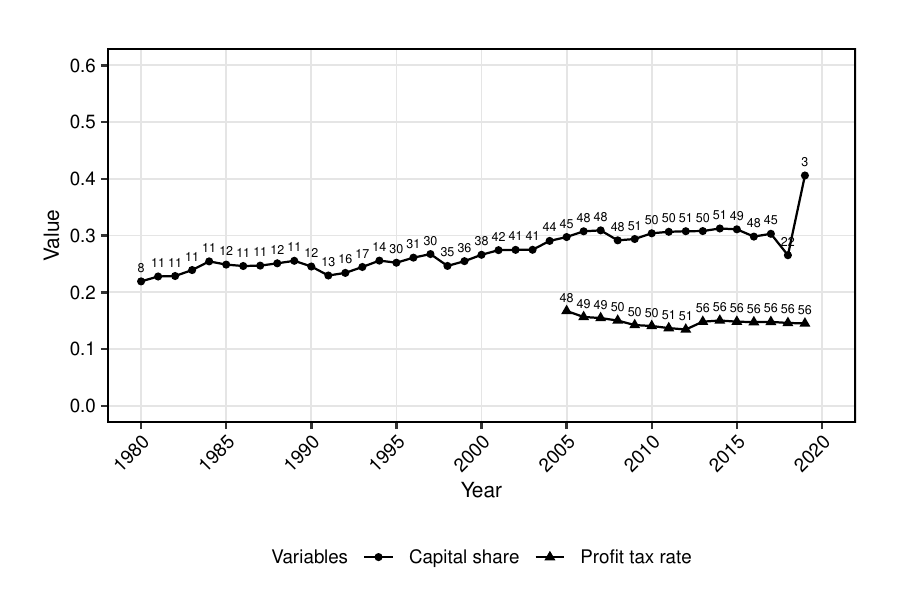}  
    \end{subfigure}
    \caption{Cross-sectional averages of the main variables in the data set}
    \label{fig:cross_averages}
\end{figure}

Fig. \ref{fig:cross_averages} provides an insight on how the capital share and income inequality that we use evolve over time.
Panel (a) displays the cross-sectional averages of top income shares whereas Panel (b) presents the cross-sectional averages of the capital share as well as the profit tax rate.
Both capital share and top income share exhibit a steady increase during roughly the first part of the period.
Average Top 5\% increases from around 20\% to about 30\% between 1980 and 2002.
After 2002, it remains stable.
Top 1\%, on the other hand, continues to rise until 2008.
It is important to note the difficulty of interpretation of the few last data points, as it seems that these jumps are related to the sharp decline in the number of countries used to calculate the averages.
Average capital share, also starting from around 20\% in 1980, reaches 30\% by 2007.
After a drop in 2008, it stabilizes around 30\%.
This first glance on the data suggests that the top income shares increase along with the capital share until the mid-2000s for Top 5\% and Top 10\%, and until 2008 for Top 1\%.
After that both capital share and top income shares remain stable.
The profit tax rate shows a steady decrease from 2005 to 2012, stabilizing after a rise in 2013.

\section{Empirical Methodology}\label{sec:empiricalmethodology}

In this section, we describe the empirical methodology that will be followed to estimate the relationship between income inequality and the capital share.
Section \ref{sec:empirical_model} introduces the empirical model based on the theoretical discussion of Section \ref{sec:theoreticalframework}.
In Section \ref{sec:identification} we discuss the identification assumptions placed on the empirical model.
Finally, Section \ref{sec:estimation} presents the two-step grouped time-varying coefficients (TVC) estimators.

\subsection{The empirical model}\label{sec:empirical_model}

In the light of the theoretical discussion of Section \ref{sec:theoreticalframework}, consider the following model of the observed top income shares:
\begin{equation}\label{eq:general_emp_model}
S_{it} = \delta_{it} \times CS_{it} + \lambda_{it} + \varepsilon_{it}, \; i=1,\dots,N, \; t=1,\dots,T,
\end{equation}
where $S_{it} = S^0_{it} + e_{y,it}$ and $CS_{it} = CS^0_{it} + e_{x,it}$
with $e_{y,it}$ and $e_{x,it}$ being $i.i.d.$ random variables capturing the deviation of the observed variables from their theoretical counterparts in Section \ref{sec:theoreticalframework},
$\delta_{it}$ and $\lambda_{it}$ are parameters to be estimated,
and $\varepsilon_{it} = -\delta_{it} e_{x,it} + e_{y,it}$ is the composite error term.
Here, as in the rest of the paper, we drop the subscript $q$ from the variables and the parameters for notational economy.

In Model \eqref{eq:general_emp_model}, the parameter $\lambda_{it}$ represents the variable $S^0_{q,L}$ of Section \ref{sec:theoreticalframework}, whereas $\delta_{it} = \kappa_{it} - \lambda_{it}$ with $\kappa_{it}$ representing $S^0_{q,K}$.
As we treat $\lambda_{it}$ as a parameter, we allow it to be arbitrarily correlated with $CS_{it}$.
However, since the composite error term includes $-\delta_{it} e_{x,it}$ as a component, it is potentially correlated with $CS_{it}$.
The identification assumptions related to this issue are discussed in Section \ref{sec:identification}.

Since the number of parameters exceed the number of observations, the model parameters are not estimable using the usual procedures such as least squares or maximum likelihood.
To obtain a model whose estimation is feasible, further constraints will be placed on the parameters.
A straightforward way to do so is to assume that the slope of the model satisfies $\delta_{it} = \delta$, and $\lambda_{it}$ is given by either $\lambda_{it} = \mu_{i}$ or $\lambda_{it} = \mu_{i} + \omega_{t}$.
The model can then be estimated using one-way or two-way fixed effects estimators, respectively.
Similar assumptions, in one or another form, have been relied on in the previous literature such as \citet{daudey2007personal}, \citet{francese2015functional}, \citet{schlenker2015capital}, \citet{bengtsson2018capital}, \citet{erauskin2020labor}.
However, as is shown in Section \ref{sec:theoreticalframework}, the coefficient of the capital share in this model, $\delta_{it}$, represents $D^0_{q}$, the difference between the top earners' share of the capital and the labor income, which is unlikely to be constant over time or for different countries.
We now discuss the assumptions we place on the time patterns and the cross-sectional heterogeneity properties of the parameters in order to obtain an estimable model.
\medskip

\textit{Time patterns of the parameters.}
Although we cannot assume that the parameters of our model are constant over time, it is also unlikely that they have very high frequency movements.
For a limited sample of countries and based on LIS surveys, the elasticity calculations provided by \citet{milanovic2017} shows that this is indeed the case.
We hence assume that the parameters of our model can be well represented by smooth functions of $t/T$.
Namely, for the coefficient of the capital share, we assume that $\delta_{it} = \delta_{i} (t/T)$ and the intercept of the model is decomposed into two terms as $\lambda_{it} = \mu_{i} + \omega_{it}$ where $\omega_{it} = \omega_{i} (t/T)$.
Here, the parameter $\mu_{i}$ is always left unrestricted to allow for country-specific shifts, in order to control for the highest amount of heterogeneity possible in the intercept.
This may also help controlling the effect of the measurement errors in the dependent variable.
\medskip

\textit{Constraints on cross-sectional heterogeneity.}
Both the literature using microdata, such as \citet{milanovic2017}, and macrodata, such as \citet{bengtsson2018capital}, show that there is considerable heterogeneity among countries in the relationship between the capital share and income inequality.
Under the smoothness assumption above, it would be possible to estimate the functions $\delta_{i} (t/T)$ separately for each country.
However, due to the scarcity of the data, especially for emerging economies, this would limit the sample to be considered.
To allow for an admissible level of cross-sectional heterogeneity of the parameters while keeping the maximum number of countries in the sample, we follow the recent literature in panel data and assume that the model parameters follow grouped time patterns.
That is, we write
\begin{equation*}
\begin{split}
    \delta_{i} (t/T) &= \sum_{g=1}^G \delta_{g} (t/T) \mathbf{1} \{ i \in \mathcal{G}_g \}, \\
    \omega_{i} (t/T) &= \sum_{g=1}^G \omega_{g} (t/T) \mathbf{1} \{ i \in \mathcal{G}_g \},
\end{split}
\end{equation*}
where $\mathcal{G}_g$, $g=1,\dots,G$ are the sets of country indexes belonging to group $g$ which satisfy $\cup_{g=1}^G \mathcal{G}_g = \{1,\dots,N\}$ and $\mathcal{G}_g \cap \mathcal{G}_{g'} = \emptyset$ for any $g \neq g'$.
One way to classify the countries in the sample would be to consider the institutional proximity between them.
\citet{bengtsson2018capital} consider Anglo-Saxon, Continental and Nordic countries as different groups, for instance.
The results of \citet{ranaldi2022capitalist} provide a very clear grouping of economies in the relationship between compositional and personal inequality.
Following the latter study, we use an estimation method which classifies the countries based on data, instead of relying on pre-determined groups.
\medskip

Under these constraints, we obtain the following model:
\begin{equation}\label{eq:grouped}
  S_{it} = \delta_{g_i} (t/T) \times CS_{it} + \mu_{i} + \omega_{g_i} (t/T) + \varepsilon_{it},
\end{equation}
where $g_i \in \{1,2,\dots,G\}$ is the group membership variable stating the group which $i$th unit belongs to.
A summary of the components of the empirical model is reported in Table \ref{tab:modelsummary}.
When $G=1$, the model reduces to a fixed effects panel data model with time-varying coefficients and it can be estimated using kernel-smoothing \citep{li2011non} or penalized splines \citep{putz2018penalized}.
Same methods can be applied when $G>1$ for pre-determined country groups $\mathcal{G}_g$, $g=1,\dots,G$.
For unknown groups, \citet{chen2019estimating} and \citet{su2019sieve} propose using kernel-smoothing and polynomial-splines combined with hierarchical clustering, and classifier-Lasso to classify the units, respectively.
In this paper, we propose a computationally simple two-step approach to estimate the unknown groups and subsequently the time-varying coefficients.

\begin{table}
  \centering
  \caption{Summary of the empirical model components}
    \begin{tabular}{ll}
    \toprule
    \multicolumn{1}{l}{Symbol} & \multicolumn{1}{l}{Definition} \\
    \midrule
    \multicolumn{2}{c}{\textit{Observables}} \\
    \midrule
    $S_{it}$     & Top income share (dependent variable) \\
    $CS_{it}$    & Capital share (explanatory variable) \\
    $PTR_{it}$    & Profit tax rate (instrument) \\
    \midrule
    \multicolumn{2}{c}{\textit{Unobservables (Treated as Parameters)}} \\
    \midrule
    $\delta_{g_it} (=\kappa_{it} - \lambda_{it})$ & Transmission coefficient \\
    $\lambda_{it} (=\mu_{i} + \omega_{g_it})$ & Labor income inequality \\
    $\kappa_{it}$ & Capital income inequality \\
    $\mu_{i}$    & Time-invariant component of labor income inequality \\
    $\omega_{g_it}$ & Time-varying component of labor income inequality \\
    \bottomrule
    \end{tabular}%
  \label{tab:modelsummary}%
\end{table}%

\subsection{Identification}\label{sec:identification}

Before proceeding with the estimation strategy, a discussion on the identification of the model parameters is needed.
First, it is easily seen that $\mu_{i}$ and the mean of $\omega_{g} (t/T)$ cannot be identified separately.
To overcome this problem, it is simply assumed that $\int_0^1 \omega_{g} (\tau) d\tau = 0$ for all $g=1,\dots,G$ \citep{hastie1986}.
Second and more importantly, the potential correlation between $\varepsilon_{it} = -\delta_{it} e_{x,it} + e_{y,it}$ and $CS_{it}$ poses a particular difficulty on the identification of the parameter of interest, $\delta_{g}(t/T)$.
It is quite reasonable to assume that the measurement error of the top income shares, $e_{y,it}$, is uncorrelated with capital share.
However, the measurement error of the capital share, $e_{x,it}$, is very likely to be correlated with the observed capital share itself.
The assumption that the measurement error is uncorrelated with the true value of the variable is usually called the \textit{classical errors-in-variables} assumption in econometrics and it causes the endogeneity of regressors \citep{wooldridge2010econometric}.
Hence, unless $\mathrm{Var}(e_{x,it}) = 0$, it is needed to instrument the capital share to be able to identify the function $\delta_{g}(t/T)$.
A potential instrument for the capital share is profit tax rate.
This is because the tax policy has a direct effect on the capital accumulation of an economy which suggests a negative relationship between the two variables.
Moreover, profit tax rate is very likely to be uncorrelated with the measurement errors in the capital share and the top income shares which makes it a good instrument.
Below, we estimate the empirical model \eqref{eq:grouped} under the assumption that $\mathrm{Var}(e_{x,it}) = 0$ as well as by instrumenting it with profit tax rate.

\subsection{Two-step Grouped TVC estimation}\label{sec:estimation}

In this section, we introduce a two-step estimation strategy following \citet{bonhomme2019distributional,bonhomme2022discretizing}.
In the first step, we classify the countries into $G$ groups via Kmeans, using the time averages of their observed variables, the top income shares and the capital shares.
We then use the estimated groups to estimate the time-varying model parameters using PLS.
Contrary to the methods proposed by \citet{chen2019estimating} and \citet{su2019sieve}, our methodology is very simple to implement using standard statistical software and it is tractable in the sense that at each step we obtain results which are interpretable from an economic perspective.

In their seminal contribution, \citet{bonhomme2019distributional} develop a two-step grouped fixed effects estimator which rely on individual-specific moments that are informative about group membership.
The assumption that the individual-specific moments can be used to learn about the group membership is called the \textit{injective moments} assumption.
In \citet{bonhomme2019distributional}, the group specific heterogeneity, whether it is time-invariant or time-varying, concerns only the intercept of their model.
In our case on the other hand, the slope parameter of the model is time-varying.
Nevertheless, the injective moments condition is easily verified noting that
$$
\bar{S}_{i} = \bar{\delta}_{g_i} \overline{CS}_{i} + \bar{\lambda}_{g_i} + \frac{1}{T} \sum_{t=1}^T \tilde{\delta}_{g_it} \widetilde{CS}_{it} + \bar{\varepsilon}_{i},
$$
where $\bar{Z}_i = T^{-1} \sum_{t=1}^T Z_{it}$ and $\widetilde{Z}_i = Z_{it} - \bar{Z}_i$ for any variable $Z_{it}$.
Under mild conditions, such as the usual mixing conditions, we have $\bar{\varepsilon}_{i} = o_p(1)$ \citep[for instance, Theorem 3.47 of][]{white01}.
Assuming also that $\overline{CS}_{i} \equiv \alpha_{g_i} + o_p(1)$ and $\plim_{T \to \infty} T^{-1} \sum_{t=1}^T \tilde{\delta}_{g_it} \widetilde{CS}_{it} \equiv \Omega_{g_i}$, for all $i=1,\dots,N$, with $\Omega_g = O(1)$, it is easily seen that one can use the time averages $\bar{S}_{i}$ and $\overline{CS}_{i}$ to classify the countries into their respective population groups.

We now describe the two steps of our estimation strategy.
We start with the classification step which uses the Kmeans estimator based on the above result on injective moments.
Second, we introduce the estimation step which uses a PLS estimator.
\medskip

\textbf{Step 1: Classification.}
Let $h_{it} = (S_{it},CS_{it})'$ and define $\bar{h}_{i} = T^{-1} \sum_{t=1}^T h_{it}$.
For a given $G$, the Kmeans estimator of the group membership variables and the group centers is defined as the solution to the following optimization problem:
\begin{equation}\label{eq:kmeans}
   \left(\widehat{\theta}_{1},\dots,\widehat{\theta}_{G},\hat{g}_1,\dots,\hat{g}_N\right) = \argmin_{\left(\theta_1,\dots,\theta_G,g_1,\dots,g_n\right)} \frac{1}{N} \sum_{i=1}^N \left\lVert \bar{h}_{i} - \theta_{g_i} \right\rVert^2,
\end{equation}
where $\theta_{g}$ is a $2 \times 1$ vector.
The global minimum of the optimization problem is not guaranteed to be achieved in finite samples but efficient algorithms to find the local minimum exist.
We use the iterative algorithm of \citet{lloyd82} for this purpose with 100 random initial values and the resulting estimates are the ones which provide the minimum over these 100 solutions to \eqref{eq:kmeans}.

The right-hand side of Equation \eqref{eq:kmeans} depends on $q$ through $S_{it} \equiv S_{q,it}$.
Here, it is implicitly assumed that the country groups do not depend on the top income share under consideration, which is a reasonable assumption.
In our implementation, we use Top 10\% to classify the countries as it contains each of Top 5\% and Top 1\% as a component.
\medskip

\textit{Choosing $G$.}
An important consideration in Kmeans estimation is the number of country groups to be used.
To choose the number of groups, we use the following Bayesian Information Criterion (BIC):
\begin{equation}\label{eq:bic}
BIC(G) = \frac{1}{N} \sum_{i=1}^N \left\lVert \bar{h}_{i} - \widehat{\theta}_{G,\hat{g}_i} \right\rVert^2 + \widehat{\sigma}^2 2G \frac{\zeta \log(N)}{N},
\end{equation}
where $\widehat{\theta}_{G,\hat{g}_i}$ being the solution to \eqref{eq:kmeans} with $G$ clusters, $\widehat{\sigma}^2$ a consistent estimate of $\mathrm{Var}(\bar{\varepsilon}_{g_it})$, and $\zeta$ is a tuning parameter.
Following \citet{lumsdaine2023estimation}, we set $\zeta=3$ in our application, and estimate the error variance using $\widehat{\sigma}^2 = [2(N-G_{max})]^{-1} \sum_{i=1}^N \lVert \bar{h}_{i} - \widehat{\theta}_{G_{max},\hat{g}_i} \rVert^2$.
The  estimate of the number of groups is then given by $\widehat{G} = \argmin_{G \in \{2,\dots,G_{max}\}} BIC(G)$.
As we treat the case of $G=1$ separately, BIC is calculated only for $G \in \{2,\dots,G_{max}\}$.
The maximum number of groups $G_{max} = 5$.
\medskip

\textbf{Step 2: Estimation.}
Having estimated the group membership variables, we estimate the following model for each estimated cluster:
\begin{equation}\label{eq:final_estimation_model}
    S_{it} = \delta_{\hat{g}_it} \times CS_{it} + \mu_{i} + \omega_{\hat{g}_it} + \varepsilon_{it},
\end{equation}
which is a panel data version of varying coefficient models first introduced by \citet{hastie93}.
To estimate the time-varying coefficients, we use polynomial splines following \citet{su2019sieve}.
Alternatively, one could use kernel-smoothing as in \citet{chen2019estimating} who uses hierarchical clustering to classify the panel units.
Due to their well documented numerical stability and good approximation properties, we focus on the B-splines in our implementation.
For details on the discussion on splines, we refer to the survey by \citet{chen2007large}.

Let $B(\tau)$, $\tau \in [0,1]$, be the $J$-dimensional B-spline basis system \citep[see, for instance, Section 4.1.4 of][]{wood2017generalized}.
Define $x_{it} = (1,CS_{it})'$ and $\gamma_{\hat{g}_i}(\tau) = [\omega_{\hat{g}_i} (\tau),\delta_{\hat{g}_i} (\tau)]'$.
The function $\gamma_{\hat{g}_i}(\tau)$ can then be approximated using $\pi_{\hat{g}_i}' B(\tau)$ where $\pi_{\hat{g}_i}$ is a $J \times 2$ matrix.
The model in \eqref{eq:final_estimation_model} can now be written as
$
    S_{it} = X_{it}'\beta_{\hat{g}_i} + \mu_{i} + U_{it},
$
where $X_{it} = [x_{it} \otimes B(t/T)]'$, $\beta_{\hat{g}_i} = \mathrm{vec}(\pi_{\hat{g}_i})$, and $U_{it} = \varepsilon_{it} + \gamma_{\hat{g}_i}(t/T)'x_{it} - X_{it}'\beta_{\hat{g}_i}$.
Then the PLS estimators of $\beta_{g}$, $g = 1,\dots,\widehat{G}$, and $\mu_{i}$ are given by:
\begin{equation}\label{eq:estimators}
\begin{split}
    \hat{\beta}_{g}(\widehat{\gamma},\hat{\psi}_g) &= \left(\widetilde{X}_g(\widehat{\gamma})'\widetilde{X}_g(\widehat{\gamma}) + \sum_{k=1}^2 \hat{\psi}_{k,g} A_{k,g}\right)^{-1} \widetilde{X}_g(\widehat{\gamma})'\widetilde{S}_{g}(\widehat{\gamma}), \\
    \hat{\mu}_{i}(\widehat{\gamma},\hat{\psi}_g) &= \frac{1}{T} \sum_{t=1}^T [S_{it}(\widehat{\gamma}) - X_{it}'\hat{\beta}_{g}(\widehat{\gamma},\hat{\psi}_g)],
\end{split}
\end{equation}
where $\hat{\psi}_g = (\hat{\psi}_{1,g},\hat{\psi}_{2,g})'$ is a vector of tuning parameters controlling the wiggleness of the estimated smooth functions, $\widetilde{X}_g(\widehat{\gamma})$ is the $\hat{n}_gT \times 2J$ data matrix stacking $\widetilde{X}_{it} = X_{it} - \bar{X}_{it}$ such that $i \in \widehat{\mathcal{G}}_g$ with $\hat{n}_g = \mathrm{Card}(\widehat{\mathcal{G}}_g)$ and $\widehat{\mathcal{G}}_g$ being the Kmeans estimator of $\mathcal{G}_g$. Similarly, $\widetilde{S}_{g}(\widehat{\gamma})$ is the $\hat{n}_gT$-dimensional data vector of the dependent variable of group $g$, and $A_{k,g}$, $k=1,2$, are matrices of known coefficients.
Let $\psi_g = (\psi_{1,g},\psi_{2,g})' \in \mathbb{R}^2$.
The penalty vector $\hat{\psi}_g$ is chosen by minimizing the generalized cross-validation score \citep[][Section 4.5.3]{wood2017generalized}:
\begin{equation*}
    \hat{\psi}_g = \argmin_{\psi_g \in \mathbb{R}^2} \frac{\hat{n}_g T \sum_{i=1}^N \sum_{t=1}^T [S_{it} - \widehat{S}_{it}(\psi_g)]^2}{[\hat{n}_g (T-1) - \sum_{k=1}^2 \mathrm{tr}(A_{k,g})]^2},
\end{equation*}
where $\widehat{S}_{it} = X_{it}'\hat{\beta}_{\hat{g}_i}(\widehat{\gamma},\psi_g) + \hat{\mu}_{i}(\widehat{\gamma},\psi_g)$ with $\hat{\beta}_{\hat{g}_i}(\widehat{\gamma},\psi_g)$ being the estimator of the form given in \eqref{eq:estimators}.

We construct the confidence bands for the time-varying coefficients and the country-specific shifts using the Bayesian posterior covariance matrix which are given by
\begin{equation*}\label{eq:variances}
\begin{split}
\widehat{\Sigma}_{g} &= \left(\widetilde{X}_g(\widehat{\gamma})'\widetilde{X}_g(\widehat{\gamma}) + \sum_{k=1}^2 \hat{\psi}_{k,g} A_{k,g}\right)^{-1} \hat{v}^2, \\
\hat{\sigma}^2_{\hat{\mu}_i} &= \left[ \frac{1}{\hat{n}_g T} + \bar{X}_{-1,g}(\widehat{\gamma})' \left(\widetilde{X}_g(\widehat{\gamma})'\widetilde{X}_g(\widehat{\gamma}) + \sum_{k=1}^2 \hat{\psi}_{k,g} A_{k,g}\right)^{-1} \bar{X}_{-1,g}(\widehat{\gamma}) \right] \hat{v}^2
\end{split}
\end{equation*}
where $\hat{v}^2 = [\hat{n}_g (T-1) - \sum_{k=1}^2 \mathrm{tr}(A_{k,g})]^{-1} \sum_{i \in \widehat{\mathcal{G}}_g} \sum_{t=1}^T \hat{\varepsilon}_{it}^2$ is an estimate of the error variance $\mathrm{Var}(\varepsilon_{it})$ and $\bar{X}_{-1,g}(\widehat{\gamma})$ is the vector of means of the right-hand side variables  $\widetilde{X}_g(\widehat{\gamma})$ excluding the intercept.
It is easily seen that, we implicitly assume within-cluster homoskedasticity and independence, although each cluster can have their own error variance.
This is clearly an unrealistic assumption.
To check the robustness of the results that we report below, we also tried frequentist variance estimation using the heteroskedasticity-consistent estimator of \citet{white1980heteroskedasticity} as well as a HAC generalization of it as in \citet{newey1987simple}.
Both estimators result in narrower confidence bands compared to the Bayesian estimate.
Hence, we choose the relatively conservative Bayesian posterior covariance matrix.
For a discussion of the differences between the frequentist and Bayesian approaches on the confidence band formation, we refer to Section 4.8 of \citet{wood2017generalized}.

\medskip
\textit{Incorporating the IV.} The literature on the estimation of time-varying coefficients panel data models using an instrumental variables approach is scarce.
Recently \citet{bai2023mean} propose a Mean Group estimator for the cross-sectional mean of the time-varying coefficient of an endogenous variable in a large panel data model.
Contrary to our case, their focus is on the average effect over all panel units.
To estimate the grouped time patterns of the effect of the potentially endogenous capital share, we use a two-stage procedure as in \citet{bai2023mean}.
Namely, we use the following projections from the reduced form model for capital share using the profit tax rate as a predictor:
$$
\widehat{CS}_{it} = \hat{\pi}_{1i} PTR_{it} + \hat{\pi}_{0i} + \hat{\pi}_{0t},
$$
where $\hat{\pi}_{1i}$, $\hat{\pi}_{0i}$ and $\hat{\pi}_{0t}$ are the linear projection coefficients and $PTR_{it}$ is the profit tax rate.
It is convenient to allow for heterogeneous coefficients in this projection step as otherwise it is possible to encounter a weak instrument problem \citep{choi2024latent,abadie2024instrumental}.
However, contrary to \citet{bai2023mean} we do not allow for time-varying coefficients here as the data on profit tax rate is short enough and heterogeneity between countries appear to be a more important issue.
Our IV estimates, called TVC-IV, are the result of Step 2 described above with $CS_{it}$ replaced with $\widehat{CS}_{it}$.

\section{Results}\label{sec:results}

In this section we present the empirical results.
We start by reporting the average effects over countries and time using the full sample, for all income shares, Top 10\%, Top 5\%  and Top 1\%.
For the time-varying coefficients and grouped estimations, we focus mainly on the results with Top 5\%.
The results for Top 10\% and Top 1\% are largely in line with the ones reported here.
The full set of results are available from the authors upon request.
Section \ref{sec:full_sample} presents the estimation results using the full sample under the assumption that $G = 1$.
In Section \ref{sec:country_groups}, we discuss the country groups resulting from the Kmeans estimator, and Section \ref{sec:grouped_results}, the estimation results using these country groups.
Finally, in Section \ref{sec:shapley}, the results on the Shapley decomposition of the predicted income inequality are reported.

\subsection{Full sample results}\label{sec:full_sample}

Before proceeding with the TVC and Grouped TVC estimation results of the effect of the capital share on income inequality, we report the results for the pooled effect over the period and countries in hand using different estimators.
These baseline results are presented in Table \ref{tab:averageeffects_fullpanel}.
The first two rows report the results of the OLS Mean Group estimation \citep{pesaran1995estimating} and the CCE Mean Group estimation \citep{pesaran06}, respectively.
Last two rows report the average effects calculated from the TVC and TVC-IV estimators described in Section \ref{sec:empirical_model} with $G = 1$.
In addition to these average effects, Table \ref{tab:averageeffects_fullpanel} reports the associated $t$-statistic in parentheses and the BIC calculated for each model in square brackets.

\begin{table}
  \caption{Estimates of the average effect of the capital share on income inequality: Full panel} 
   \label{tab:averageeffects_fullpanel}%
  \centering
  \begin{threeparttable}
    \begin{tabular}{p{2.5cm}p{2.5cm}p{2.5cm}p{2.5cm}}
    \toprule
    Estimator & Top 10\% & Top 5\% & Top 1\% \\
    \midrule
    OLS   & 0.44  & 0.46  & 0.38 \\
          & (3.92) & (3.44) & (3.05) \\
          & [-6.3] & [-6.29] & [-6.6] \\
          &       &       &  \\
    CCE   & 0.19  & 0.25  & 0.1 \\
          & (1.98) & (2.2) & (1.17) \\
          & [-5.07] & [-5]  & [-5.21] \\
          &       &       &  \\
    TVC   & 0.11  & 0.17  & 0.14 \\
          & (4.48) & (4.75) & (5.09) \\
          & [-6.96*] & [-6.96*] & [-7.29*] \\
          &       &       &  \\
    TVC-IV & 0.28  & 0.28  & 0.23 \\
          & (5.63) & (5.59) & (5.59) \\
          & [-6.54] & [-6.54] & [-6.95] \\
    \bottomrule
    \end{tabular}%
    \begin{tablenotes}[para,flushleft]
    \vspace{3mm}
     \textit{Note:} The row ``OLS" reports the Mean Group estimates of the coefficients $\delta_{i}$ associated with the model $S_{it} = \delta_{i} CS_{it} + \lambda_{i} + \varepsilon_{it}$, estimated by OLS whereas the row ``CCE" contains those associated with $S_{it} = \delta_{i} CS_{it} + l_i^{\prime} f_t + \varepsilon_{it}$.
     The rows ``TVC" and ``TVC-IV" report the average effect calculated from the model and estimators described in Section \ref{sec:empiricalmethodology} with $G=1$.
     $t$-statistics for the significance of each average effect are in parentheses and the BIC are reported in square brackets.
     Smallest BIC per column is shown with an asterisk.
     The variance estimates used for OLS and CCE are robust to arbitrary heteroskedasticity, serial correlation and weak cross-dependence, and TVC and TVC-IV tests are based on the Bayesian posterior covariance matrix.
\end{tablenotes}
    \end{threeparttable}
\end{table}

The average effect is positive for all estimators and top income shares, and smaller than 1 in line with the theoretical expectations.
The OLS estimator shows relatively high effects, 0.44, 0.46 and 0.38 for Top 10\%, Top 5\%, and Top 1\%, respectively.
CCE estimator results in lower coefficients indicating that accounting for cross-sectional dependence reduces the overall estimated effect.
This suggests that controlling for unobserved common factors via CCE estimator can correct for potential endogeneity issues in Equation \eqref{eq:general_emp_model}. 
Specifically, if the capital share is correlated with labor income inequality positively, as might occur if labor and capital market liberalization policies are implemented in different countries simultaneously, then capital share could be correlated with the error term in Equation \eqref{eq:general_emp_model}.
This correlation would lead to an upward bias in the estimation of the parameter $\delta_i := \delta_{it}$ of these models.
Controlling for unobserved common factors, CCE estimator may help mitigate this bias to some extent, provided that these factors capture some of the variation in labor income inequality which is correlated with the capital share. 

The TVC estimator, which is the method chosen by the BIC, produces even lower effects, suggesting that the effect of the capital share on income inequality varies over time in a way that when computed on average, it reduces the overall effect compared to the estimators which do not account for time-varying coefficients.
Lastly, the TVC-IV estimator shows moderate effects, which are found as 0.28 for Top 10\% and Top 5\%, and 0.23 for Top 1\%.
We should note that the average effects calculated using the TVC and TVC-IV estimators are not directly comparable as the latter is estimated using only the last 16 years of the original sample due to limited time span of the instrument, the profit tax rate.

As a final remark, it is useful to compare our estimates with those found in the previous literature.
First, \citet{francese2015functional} analyzes the relationship between factor shares and the Gini coefficient, making a direct comparison of our estimates and theirs infeasible.
Notably, their estimates are very small and statistically insignificant contrary to our economically and statistically significant findings.
Second and more importantly, a comparison should be made with \citet{bengtsson2018capital} who estimate the elasticity of Top 1\% income share with respect to the capital share.
Using the full sample which includes 21 countries and the period closest to ours, namely their ``1980-present" sample, the elasticity estimate is equal to 0.51 for Top 1\% (see their Table 2, Column 4).
To make the comparison possible, we calculate the elasticity of income inequality with respect to the capital share using our marginal effect estimates.
This requires the specific values taken by the two variables because $\epsilon_{xy} = (dx/x)/(dy/y)$, hence, $dx/dy=\epsilon_{xy}x/y$ where $\epsilon_{xy}$ is the elasticity of the variable $x$ with respect to $y$.
Evaluating this equation using the average values in our sample, we find an effect equivalent to 0.22 which is lower than our OLS estimate (0.38) for the same inequality measure, Top 1\%. 
This is in line with the findings of \citet{flores2021capital} which suggest that fiscal data tend to underestimate inequality levels as well as their sensitivity to the capital share, whereas inequality series constructed using the DINA methodology highlight a greater impact of the capital share on inequality.
The difficulty in making these values directly comparable though is the differences in samples, as the estimates by \citet{bengtsson2018capital} cover 21 mostly advanced countries, while our full sample includes approximately three times as many countries, also covering emerging economies. 

\begin{figure}
    \centering
    \begin{subfigure}{0.45\textwidth}  
        \caption{Time pattern of $\hat{\delta}_t$}  
        \includegraphics[width=\textwidth]{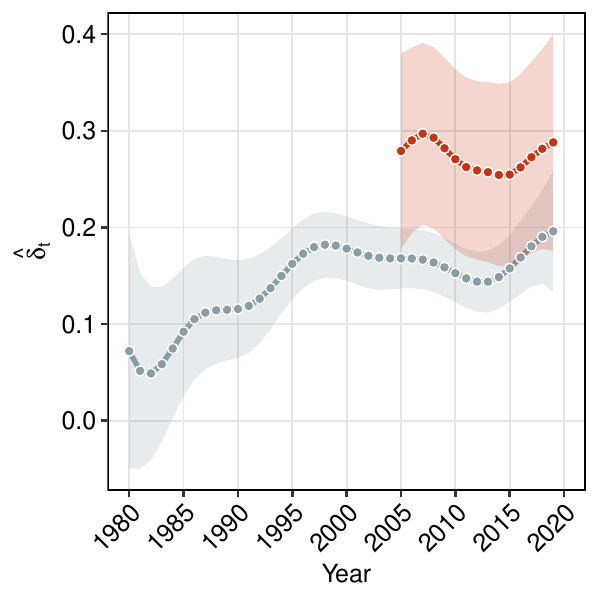}  
    \end{subfigure}
    \quad
    \begin{subfigure}{0.45\textwidth}  
        \caption{Time pattern of $\hat{\omega}_t$}  
        \includegraphics[width=\textwidth]{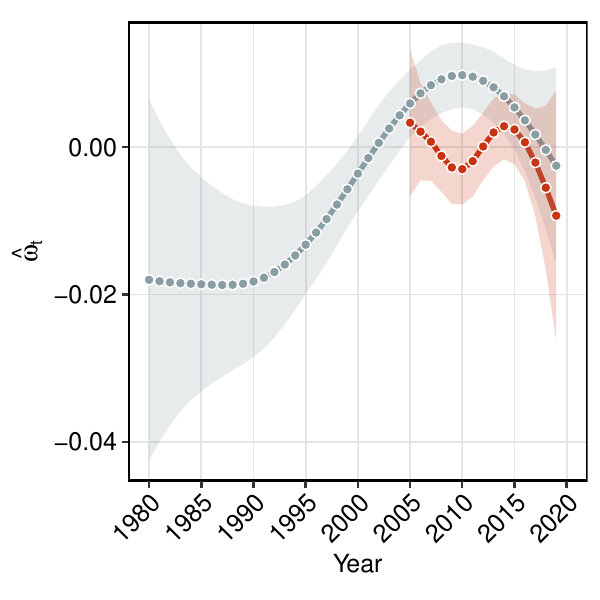}  
    \end{subfigure}
    \begin{subfigure}{0.9\textwidth}  
        \caption{Estimated country-specific shifts $\hat{\mu}_i$}  
        \includegraphics[width=\textwidth]{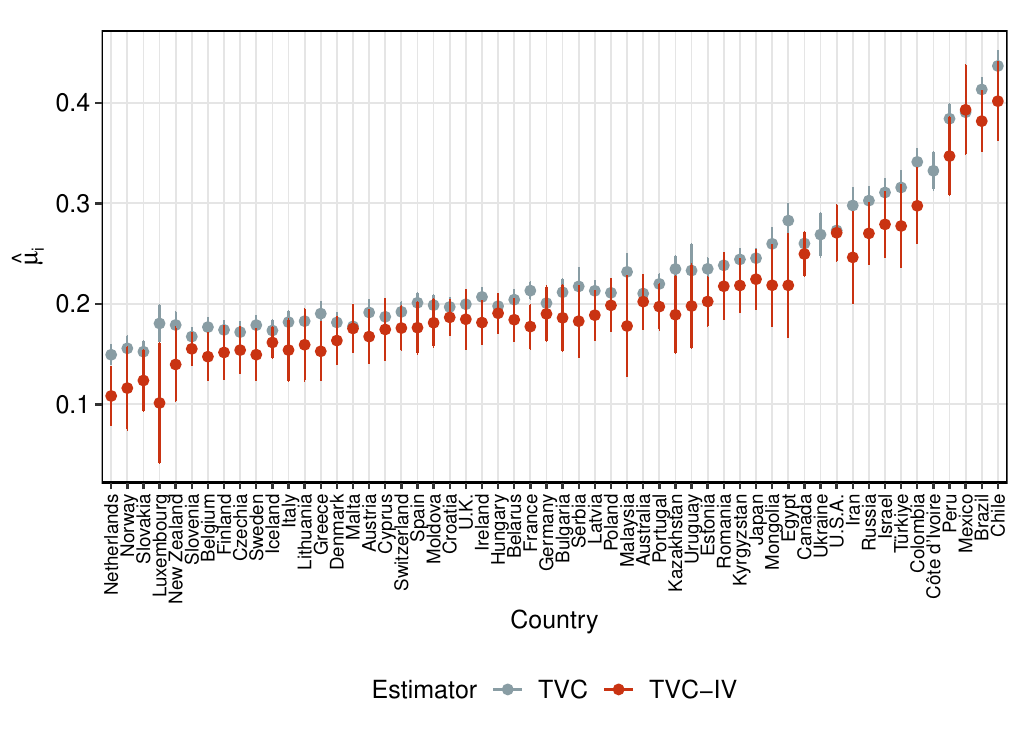}  
    \end{subfigure}
    \caption{TVC and TVC-IV estimates of the model parameters using full sample}
     \floatfoot{\footnotesize{\textit{Note:} Parameters are estimated using PLS using the Top 5\% income share as the dependent variable. 95\% confidence bands based on Bayesian posterior covariance matrix estimates are shown together with the point estimates.}}
    \label{fig:fullsample_top5}
\end{figure}

Although these first results give an idea on the average effects using different empirical model specifications, the main interest of our study is the time pattern of the transmission effect.
Fig. \ref{fig:fullsample_top5} presents the TVC (gray) and the TVC-IV (red) estimates of the model parameters.
We will interpret the results with our full sample TVC first, then comment on TVC-IV.

As is seen in Panel (a) of Fig. \ref{fig:fullsample_top5}, between 1980 and 2000, the transmission coefficient $\hat{\delta}_t$ increases significantly by around 10 pp, indicating that the impact of the capital share on inequality became stronger.
Meanwhile labor income inequality, $\hat{\omega}_t$, remains relatively stable, with a slight increase of 1.8 pp from 1990 to 2000, adding a small dampening effect on $\hat{\delta}_t$.
This suggests that the significant increase in $\hat{\delta}_t$ in the first half is due to an increase in capital income inequality $\hat{\kappa}_t$.
Interpreting this together with the simultaneous increase in capital share and income inequality from 1980 to 2000 in Fig. \ref{fig:cross_averages}, increasing capital income inequality appears to amplify the effect of capital share on inequality.
While labor income inequality $\hat{\omega}_t$ rises slightly in the 1990s, it can not offset the impact of $\hat{\kappa}_t$ on the transmission coefficient $\hat{\delta}_t$.
Moreover, it positively contributes through its direct link to inequality.
From 2000 to 2012, $\hat{\delta}_t$ decreases by around 3 pp while $\hat{\omega}_t$ rises slightly, accounting for around half of the drop in $\hat{\delta}_t$.
After 2012, $\hat{\delta}_t$ increases by about 5 pp while $\hat{\omega}_t$ returns to its 2000 level.
This suggests that $\hat{\kappa}_t$ first declines through 2008, then rises.
Incorporating this into the picture in Fig. \ref{fig:fullsample_top5} where the increase in the capital share from 2000 to 2008 is not accompanied by an increase in overall inequality, we can argue that while a growing capital share would typically drive inequality higher, this effect appears to be balanced by a decrease in the transmission coefficient.
From 2008 to 2020, both capital share and inequality remains constant, while the impact of capital income, namely $\hat{\kappa}_t$, and labor income, namely $\hat{\omega}_t$, on overall income inequality appear to balance out.

The results for the TVC-IV are mostly in line with those of the TVC except that the TVC-IV consistently provides a larger effect.
This finding is in line with the textbook result on the bias of the least squares estimates under measurement error in an explanatory variable.
As the average effect of the capital share on income inequality is expected to be positive, it is also expected that the TVC estimator underestimates the effect compared to the TVC-IV if the capital share is measured with error \citep[see, for instance,][p.80]{wooldridge2010econometric}.
TVC-IV, in principle, corrects this and leads to higher estimates.
We note that, for each of the three dependent variables, the BIC results show the superiority of the TVC estimates.
However, since the samples considered by the two estimators are dramatically different, this comparison should be interpreted with caution.

This analysis provides us a insightful overall picture regarding the pattern of the drivers of inequality.
Nevertheless, due potentially complex and heterogeneous group dynamics, the effects of different country groups may offset each other.
This points to the importance of accounting for group heterogeneity and time-varying effects together in our analysis, which is a key focus of our exercise.

\subsection{Country classification}\label{sec:country_groups}

\begin{figure}
    \centering
    \begin{subfigure}{0.6\textwidth}  
        \caption{Top income and capital share groups}  
        \includegraphics[width=\textwidth]{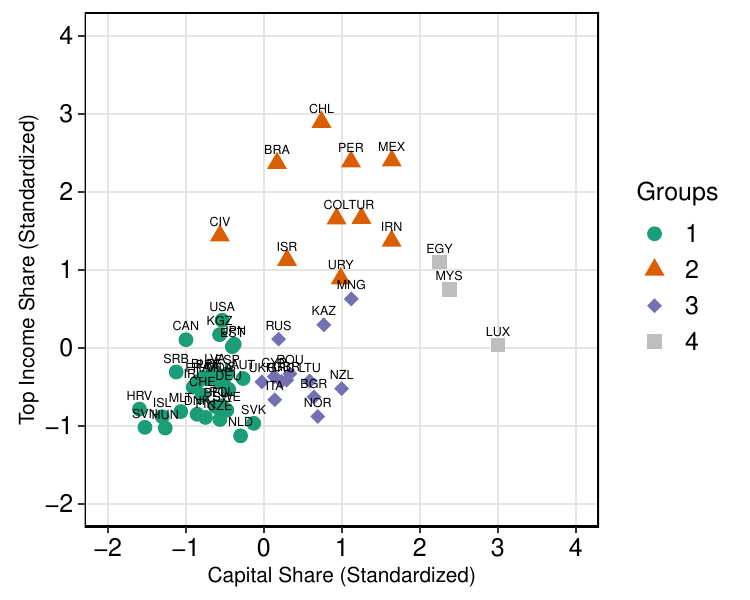}  
    \end{subfigure}
    ~
    \par\bigskip
    \begin{subfigure}{0.5\textwidth}  
        \caption{Grouped patterns of capital share}  
        \includegraphics[width=\textwidth]{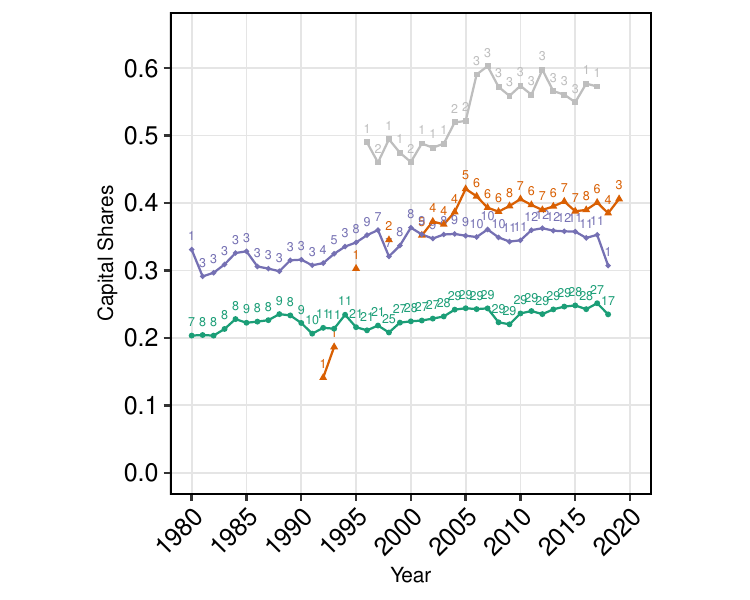}  
    \end{subfigure}
    \hspace{-1cm}
    \begin{subfigure}{0.5\textwidth}  
        \caption{Grouped patterns of top incomes}  
        \includegraphics[width=\textwidth]{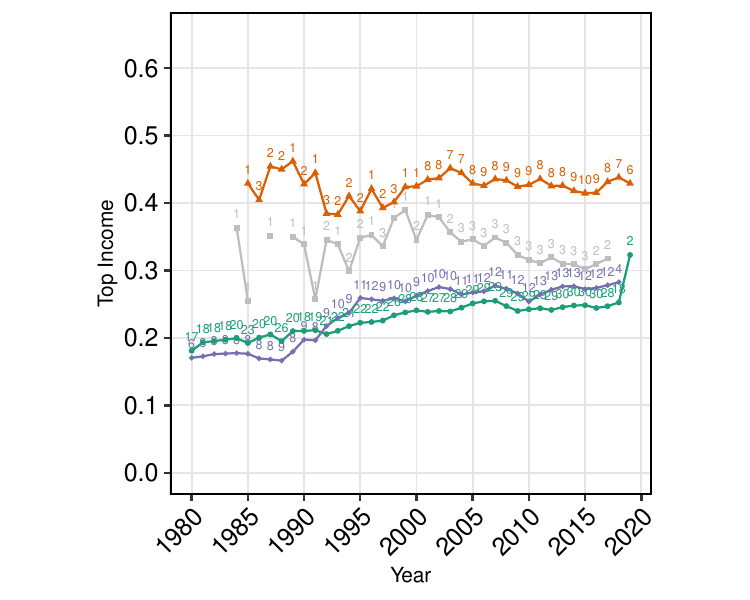}  
    \end{subfigure}
    \caption{Period averages of capital shares and Top 5\% income share (a), cross-sectional average of capital shares (b) and Top 5\% income share (c) by groups. Number of observations used for the calculation of the cross-sectional averages are shown alongside the lines in Panel (b) and Panel (c).}
    \label{fig:aosis_grouped_patters}
\end{figure}

Fig. \ref{fig:aosis_grouped_patters} presents the results of the first step of our estimation approach.
In Panel (a) of the figure, the scatter plot between the scaled and centered capital shares and top income share is reported using the time averages of the variables calculated for each country.
The scatter plot reveals a strong positive association between the time averages of the two variables.
The estimated regression line from these time averages suggests that a one standard deviation increase in the capital share is associated with an approximately 0.87 standard deviation increase in the Top 5\%.

\captionsetup[subfigure]{oneside,margin={0cm,0cm}}

\captionsetup[subfigure]{oneside,margin={1.4cm,0cm}}

The Kmeans estimates of country groups are shown in different colors.
Based on the BIC in \eqref{eq:bic}, the number of clusters is selected as $\widehat{G}=4$.
Countries with relatively low capital share and low income inequality belong to Group 1 (green) which are typically advanced economies with the addition of five transition economies (Belarus, Kyrgyzstan, Moldova, Latvia, and Serbia).
Group 2 (orange) consists of emerging economies primarily from Latin America, as well as Israel, all exhibiting high capital share and high income inequality.
Group 3 (purple) includes a mix of developed (Norway, the U.K., New Zealand, and Italy), transition (Bulgaria, Romania, Lithuania, and Ukraine), and resource-driven economies (Russia, Kazakhstan, and Mongolia) with moderate capital share but varying levels of income inequality with some countries with relatively high inequality.
However, when the group mean is considered, we can see that their level of inequality is moderate as well.
Group 4 (gray) contains three structurally different countries (Egypt, Malaysia, Luxembourg) characterized by very high capital share and high income inequality.
This group displays less informative grouped patterns compared to the other groups, making it less relevant for further analysis.
Moreover, due to the very limited number of observations, it was impossible to make meaningful TVC estimations for this group.
Thus, we treat them as outliers for the rest of the analysis.

The grouped cross-sectional averages for the capital share and Top 5\% are reported in Panel (b) and Panel (c), respectively.
Each data point in these figures is shown along with the number of countries used in the calculation of the grouped cross-sectional averages.
Group 1 shows a consistent increase in both capital share and income inequality over the entire period with small fluctuations.
Group 2 spans a shorter period and displays a more volatile pattern.
For this group, capital share first increases between 2000 and 2005, then gradually decreases until 2008.
After 2008 it stabilizes at around 0.4 with fluctuations until the end of the period.
Income inequality, on the other hand, rises between 1997 and 2003, followed by gradual a decline until 2016.
Group 3 shows a stable picture of income inequality until the late 1980s while the capital share fluctuates around 0.3.
This is followed by a sharp increase in both capital share and top incomes until 1997 and 1995, respectively, surpassing the average of Group 1 in income inequality.
This is followed by a fall in capital share in 1997, which quickly rises back to the pre-fall levels by 2000, while no change in income inequality is observed between 1995-2000.
From 2000 onward, both capital share and income inequality remains stable with minor fluctuations around 0.35 and 0.28, respectively.

\subsection{Grouped patterns}\label{sec:grouped_results}

Before proceeding with the grouped time patterns, it will prove useful to have a glance at the average effects by group.
Table \ref{tab:grouped_averageeffects} shows the group-specific average effects of the capital share on Top 5\%, also estimated using OLS, CCE, TVC, and TVC-IV.
For Group 1, the OLS estimate of 0.41 shows a moderate effect.
As in the case of the full sample, CCE provides a weaker effect of 0.34, confirming that controlling for common factors reduce the estimated effect of the capital share on income inequality.
TVC and TVC-IV estimates (0.20 and 0.29, respectively) indicate that accounting for time variation leads to a further reduction in the effect, with the capital share still playing a significant role in driving income inequality.
For Group 2, the OLS estimate is found to be 1.11 which is unusually high.
This suggests a very strong effect of the capital share, but CCE and TVC show much lower effects (0.30 and 0.22, respectively).
TVC-IV increases the effect to 0.44, probably reflecting the growing importance of capital income inequality in emerging markets in the last years of the sample.
For Group 3, the OLS shows a relatively low effect of 0.22, and CCE provides a slightly higher estimate which is equal to 0.25.
TVC and TVC-IV provide more moderate effects (0.16 and 0.15, respectively), indicating that the capital share plays a smaller but consistent role in these diverse countries with fluctuating inequality patterns.
Overall, the results show significant heterogeneity across groups.
It is seen that the capital share has a stronger average effect in countries with high capital share and high income inequality which are mostly emerging countries.
Whereas it has a moderate average effect in countries with low capital share and low income inequality which are mainly advanced economies.
The role of the capital share is found to be less pronounced but still relevant in the the country group with moderate capital share levels and moderate income inequality.

\begin{table}
  \caption{Estimates of the average effect of the capital share on Top 5\%: Grouped estimates} 
   \label{tab:grouped_averageeffects}%
  \centering
  \begin{threeparttable}
    \begin{tabular}{p{2.5cm}p{3.6cm}p{3.6cm}p{3.6cm}H}
    \toprule
    Estimator & Group 1 & Group 2 & Group 3 & Group 4 \\
    \midrule
    OLS   & 0.41  & 1.11  & 0.22  & 0.22 \\
          & (5.98) & (1.40) & (2.88) & (0.76) \\
          & [-7.17*] & [-5.39] & [-6.12] & [-6.59] \\
          &       &       &       &  \\
    CCE   & 0.34  & 0.30 & 0.25  & 0.07 \\
          & (5.72) & (1.79) & (3.77) & (0.25) \\
          & [-6.4] & [-3.50] & [-5.54] & [-6.39] \\
          &       &       &       &  \\
    TVC   & 0.2   & 0.22  & 0.16  & 0.01 \\
          & (11.6) & (1.8) & (4.73) & (0.13) \\
          & [-5.66] & [-6.88*] & [-6.59*] & [-7.48*] \\
          &       &       &       &  \\
    TVC-IV & 0.29  & 0.44  & 0.15  & 0.98 \\
          & (5.29) & (2.73) & (1.3) & (4.29) \\
          & [-5.32] & [-6.06] & [-3.38] & [-7.12] \\
    \bottomrule
    \end{tabular}%
    \begin{tablenotes}[para,flushleft]
    \vspace{3mm}
     \textit{Note:} See Table \ref{tab:averageeffects_fullpanel} notes. The group members are reported in Fig. \ref{fig:aosis_grouped_patters} and also in Table \ref{tab:shapley}.
    \end{tablenotes}
    \end{threeparttable}
\end{table}

\begin{figure}
    \centering
    \begin{subfigure}{0.45\textwidth}  
        \caption{Time pattern of $\hat{\delta}_{1t}$}  
        \includegraphics[width=\textwidth]{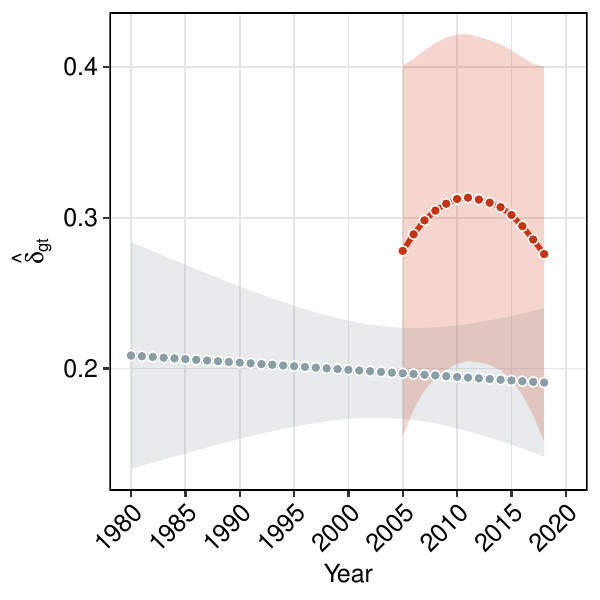}  
    \end{subfigure}
    \quad
    \begin{subfigure}{0.45\textwidth}  
        \caption{Time pattern of $\hat{\omega}_{1t}$}  
        \includegraphics[width=\textwidth]{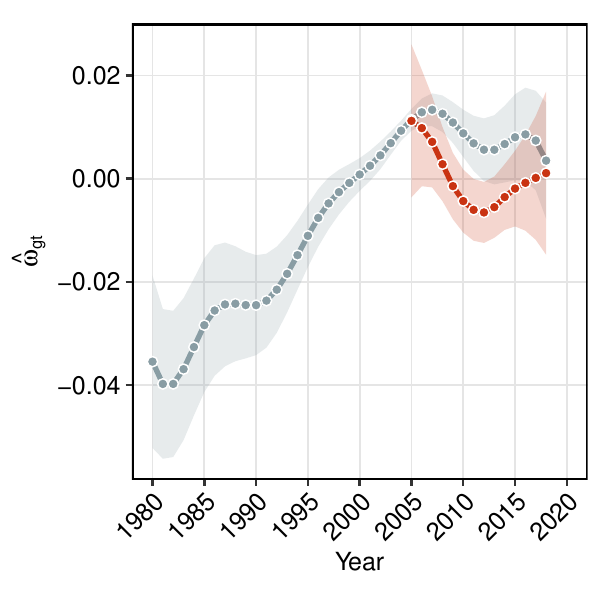}  
    \end{subfigure}
    \caption{TVC and TVC-IV estimates of the model parameters of Group 1}
     \floatfoot{\footnotesize{\textit{Note:} See Fig. \ref{fig:fullsample_top5} notes.}}
    \label{fig:group1_top5}
\end{figure}

The time patterns of the TVC estimates for Group 1  are shown in Fig. \ref{fig:group1_top5}.
We omit the results on the estimated country-specific shifts, as in the rest of the paper, which are available from the authors upon request.
In Panel (a), we observe almost no change in the transmission coefficient $\hat{\delta}_{1t}$, indicating that capital income inequality $\hat{\kappa}_{1t}$ must have followed a similar pattern to that of $\hat{\omega}_{1t}$, hence $\hat{\lambda}_{1t}$.
According to Panel (b), labor income inequality increased significantly, around 5 pp, from the beginning of the period until 2008, followed by a gradual decrease toward the end of the sample (around 1.8 pp).
Interpreting these results alongside the observed steady increase in capital share and income inequality until 2008 in Fig. \ref{fig:aosis_grouped_patters}, we conclude that while the 5 pp increase in the capital share translates into a 1 pp increase in the top income share, the remaining rise in the top income share is predominantly driven by increases in capital and labor income inequalities during this period.
After 2008, income inequality stabilizes while the capital share continues to increase, suggesting that the inequality driven by the rise in the capital share is counterbalanced by the decrease in both labor and capital income inequality ($\hat{\lambda}_{1t}$ and $\hat{\kappa}_{1t}$), maintaining the steady path of $\hat{\delta}_{1t}$.

\begin{figure}
    \centering
    \begin{subfigure}{0.45\textwidth}  
        \caption{Time pattern of $\hat{\delta}_{2t}$}  
        \includegraphics[width=\textwidth]{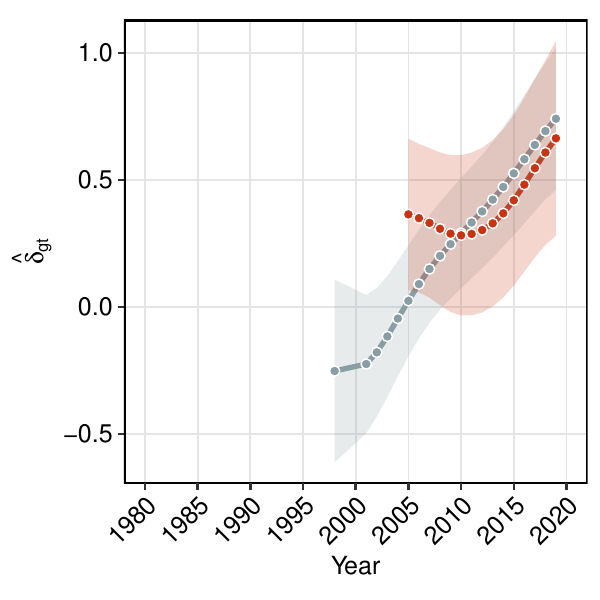}  
    \end{subfigure}
    \quad
    \begin{subfigure}{0.45\textwidth}  
        \caption{Time pattern of $\hat{\omega}_{2t}$}  
        \includegraphics[width=\textwidth]{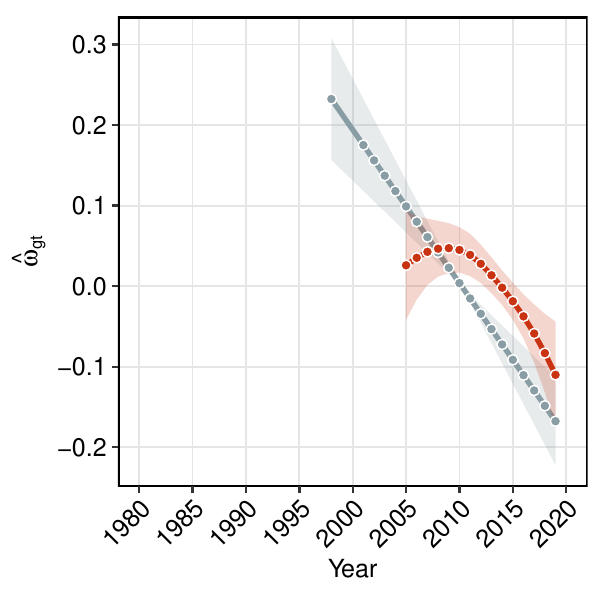}  
    \end{subfigure}
    \caption{TVC and TVC-IV estimates of the model parameters of Group 2}
     \floatfoot{\footnotesize{\textit{Note:} See Fig. \ref{fig:fullsample_top5} notes.}}
    \label{fig:group2_top5}
\end{figure}

Fig. \ref{fig:group2_top5} shows the time patterns of the model parameters for Group 2.
It is seen that, over the past two decades, $\hat{\omega}_{2t}$ has decreased significantly, which is mirrored by an increase in $\hat{\delta}_{2t}$, as confirmed by both TVC and TVC-IV estimators.
This suggests that the impact of the capital share on income inequality has significantly increased during this period, while labor income inequality has decreased.
The magnitude of the increase in $\hat{\delta}_{2t}$, however, appears to surpass the decrease in $\hat{\omega}_{2t}$, meaning that capital income inequality ($\hat{\kappa}_{2t}$), must also be rising.
Interpreting these results alongside the income inequality and capital share dynamics from Fig. \ref{fig:aosis_grouped_patters}, where both capital share and inequality seem to remain stable from 2000 onward, we can argue that the fluctuations in inequality are likely driven by a combination of factors, such as rising capital income inequality, decreasing labor income inequality, and the increase in the transmission coefficient, which converts fluctuations in capital share into overall income inequality. A formal analysis of this claim will be the topic of the next section where we apply a Shapley decomposition of the model predictions.

\begin{figure}
    \centering
    \begin{subfigure}{0.45\textwidth}  
        \caption{Time pattern of $\hat{\delta}_{3t}$}  
        \includegraphics[width=\textwidth]{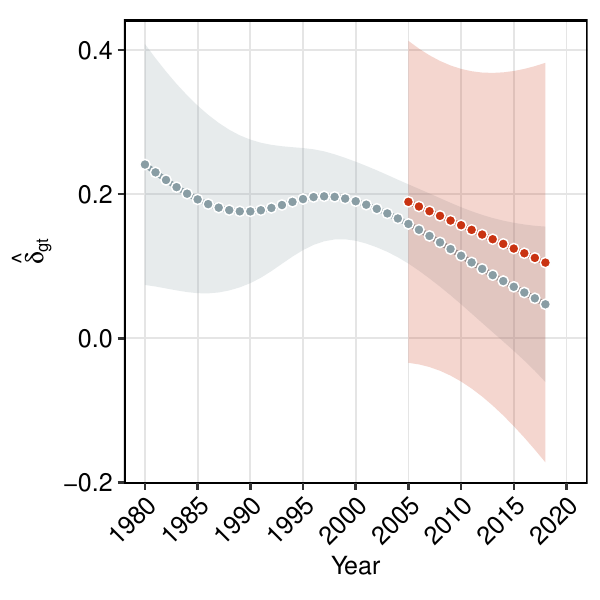}  
    \end{subfigure}
    \quad
    \begin{subfigure}{0.45\textwidth}  
        \caption{Time pattern of $\hat{\omega}_{3t}$}  
        \includegraphics[width=\textwidth]{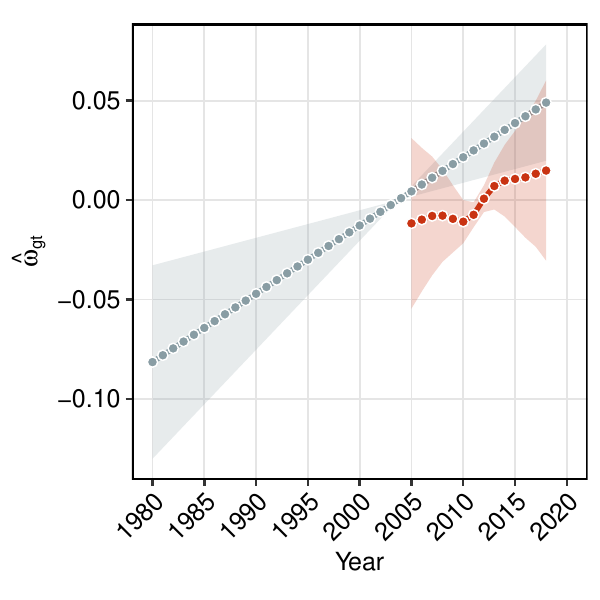}  
    \end{subfigure}
    \caption{TVC and TVC-IV estimates of the model parameters of Group 3}
     \floatfoot{\footnotesize{\textit{Note:} See Fig. \ref{fig:fullsample_top5} notes.}}
    \label{fig:group3_top5}
\end{figure}

The results on the time patterns of Group 3 are reported in Fig. \ref{fig:group3_top5}. Between the late 1980s and 2000, $\hat{\delta}_{3t}$ remains relatively stable, while $\hat{\omega}_{3t}$ shows a steady increase, suggesting that $\hat{\kappa}_{3t}$ fluctuates to account for the remaining variations in $\hat{\delta}_{3t}$. After 2000, $\hat{\delta}_{3t}$ gradually declines, higher in magnitude compared to the rise in $\hat{\omega}_{3t}$, implying a decline in $\hat{\kappa}_{3t}$ during this period.  
When observed together with the movements in capital share and income inequality from Fig. \ref{fig:aosis_grouped_patters}, the sharp rise in overall income inequality from the late 1980s to the mid-1990s appears to be driven by a rapid increase in capital share, transmitted through a stable transmission coefficient ($\hat{\delta}_{3t}$) together with a rise in labor income inequality ($\hat{\omega}_{3t}$).
After 2000, the capital income inequality decreases and labor income inequality continues to rise resulting in a decline of the transmission coefficient ($\hat{\delta}_{3t}$) which reduces the influence of the capital share on income inequality.
Together, these shifts help explain the relative stabilization of income inequality over the past two decades, despite ongoing fluctuations in its underlying components.

It is important to note that although these estimation results provide a clear exposition of time patterns of each component of the empirical model, it is still difficult to understand the contribution of each component on income inequality separately.
This is because the transmission coefficient, labor income inequality and the capital share are clearly correlated, making it difficult to separate their role in driving income inequality without additional analysis.
In the next section we provide a formal examination of these roles through a Shapley decomposition of the model predictions.

We close this section with a summary of the findings. 
Overall, the effect of the capital share on income inequality remains stable in Group 1, rises in Group 2, and declines in Group 3 over the observed period.
The labor income inequality increases for all countries except in the group of emerging countries and Israel (Group 2).
In Group 1, which is composed of mostly developed economies with low inequality and capital shares, capital and labor income inequality jointly increase until 2008.

\subsection{Decomposing the predicted income inequality}\label{sec:shapley}

In the previous sections, we discussed the estimation of the parameters of different models of income inequality.
Especially in the models where the parameters change between countries and over time, it is important to put the estimation results into a historical perspective.
By interpreting the parameter estimates together with the average changes in the sample documented in Fig. \ref{fig:cross_averages} and Fig.  \ref{fig:aosis_grouped_patters}, we tried to understand the role of each factor on the changes in income inequality.
In this section, we put the estimated models into perspective in a formal manner.
Namely, we decompose the predicted income inequality by Model \eqref{eq:final_estimation_model} for each country in the period under consideration into three components: the transmission coefficient $\delta_{gt}$, the capital share $CS_{it}$ and the time-varying component of labor income inequality $\omega_{gt}$.

Different ways to decompose the relative contribution of each explanatory variable to the $R^2$ of an income inequality regression are well-studied in the literature \citep[see, for instance,][]{shorrocks1982inequality,shorrocks1984inequality,sastre2002shapley,fields2003accounting}.
Following \citet{israeli2007shapley}, we calculate the Shapley contribution of each component of the empirical model to income inequality.
However, our model and the estimation method requires a careful reconsideration of the Shapley decomposition proposed by the author.
First, the aforementioned literature, \citet{israeli2007shapley} in particular, considers the decomposition of the $R^2$ by re-estimating the model using each subset of the full set of explanatory variables.
As the decrease in the $R^2$ following the exclusion of a particular variable from the models using different subsets of the explanatory variables represents the marginal contribution of that variable, one can calculate the average contribution of that variable.
Our estimation method, however, is not well suited to this method.
This is because we choose the smoothing parameters for the time-varying coefficients using the data in hand which in turn implies that the exclusion of a particular variable does not necessarily decrease the $R^2$ of the model.
Hence, the Shapley decomposition proposed by \citet{israeli2007shapley} are not suitable for our model and the estimation method.

\begin{table}
  \centering
  \caption{Proportion of the contribution of each component of Model \eqref{eq:final_estimation_model} to the predicted income inequality for each country in the last four decades}
    \resizebox{\textwidth}{!}{  
  \begin{threeparttable}
    \begin{tabular}{lrrrrrrrrrrrrr}
    \toprule
    \multicolumn{4}{c}{Group 1}   &       & \multicolumn{4}{c}{Group 2}   &       & \multicolumn{4}{c}{Group 3} \\
\cmidrule{1-4}\cmidrule{6-9}\cmidrule{11-14}    \multicolumn{1}{c}{Country} & \multicolumn{1}{c}{\textbf{$\delta_{
1t}$}} & \multicolumn{1}{c}{\textbf{$CS_{it}$}} & \multicolumn{1}{c}{\textbf{$\omega_{1t}$}} &        & \multicolumn{1}{c}{Country} & \multicolumn{1}{c}{\textbf{$\delta_{2t}$}} & \multicolumn{1}{c}{\textbf{$CS_{it}$}} & \multicolumn{1}{c}{\textbf{$\omega_{2t}$}} &       & \multicolumn{1}{c}{Country} & \multicolumn{1}{c}{\textbf{$\delta_{3t}$}} & \multicolumn{1}{c}{\textbf{$CS_{it}$}} & \multicolumn{1}{c}{\textbf{$\omega_{3t}$}} \\
    \midrule
    Australia & 0.27  & 0.53  & 0.20  &       & \multicolumn{1}{l}{Brazil} & 0.27  & 0.45  & 0.28  &       & \multicolumn{1}{l}{Bulgaria} & 0.21  & 0.51  & 0.27 \\
    Austria & 0.22  & 0.52  & 0.26  &       & \multicolumn{1}{l}{Chile} & 0.02  & 0.46  & 0.53  &       & \multicolumn{1}{l}{Cyprus} & 0.16  & 0.50  & 0.34 \\
    Belgium & 0.25  & 0.54  & 0.21  &       & \multicolumn{1}{l}{Colombia} & 0.29  & 0.50  & 0.20  &       & \multicolumn{1}{l}{U.K.} & 0.30  & 0.55  & 0.15 \\
    Belarus & 0.25  & 0.49  & 0.26  &       & \multicolumn{1}{l}{Iran} & 0.31  & 0.56  & 0.14  &       & \multicolumn{1}{l}{Greece} & 0.22  & 0.51  & 0.27 \\
    Canada & 0.30  & 0.47  & 0.24  &       & \multicolumn{1}{l}{Israel} & 0.33  & 0.48  & 0.19  &       & \multicolumn{1}{l}{Italy} & 0.32  & 0.57  & 0.11 \\
    Switzerland & 0.25  & 0.50  & 0.26  &       & \multicolumn{1}{l}{Mexico} & 0.30  & 0.51  & 0.19  &       & \multicolumn{1}{l}{Kazakhstan} & 0.20  & 0.51  & 0.30 \\
    Czechia & 0.23  & 0.53  & 0.24  &       & \multicolumn{1}{l}{Peru} & 0.52  & 0.48  & 0.00  &       & \multicolumn{1}{l}{Lithuania} & 0.20  & 0.54  & 0.26 \\
    Germany & 0.25  & 0.52  & 0.23  &       & \multicolumn{1}{l}{Türkiye} & 0.57  & 0.52  & -0.09 &       & \multicolumn{1}{l}{Mongolia} & 0.14  & 0.50  & 0.36 \\
    Denmark & 0.24  & 0.50  & 0.26  &       & \multicolumn{1}{l}{Uruguay} & 0.75  & 0.58  & -0.33 &       & \multicolumn{1}{l}{Norway} & 0.27  & 0.64  & 0.09 \\
    Spain & 0.24  & 0.51  & 0.25  &       &       &       &       &       &       & \multicolumn{1}{l}{New Zealand} & 0.16  & 0.56  & 0.28 \\
    Estonia & 0.25  & 0.49  & 0.27  &       &       &       &       &       &       & \multicolumn{1}{l}{Romania} & 0.23  & 0.48  & 0.29 \\
    Finland & 0.27  & 0.56  & 0.18  &       &       &       &       &       &       & \multicolumn{1}{l}{Russia} & 0.19  & 0.44  & 0.38 \\
    France & 0.28  & 0.51  & 0.21  &       &       &       &       &       &       & \multicolumn{1}{l}{Ukraine} & 0.36  & 0.50  & 0.13 \\
    Croatia & 0.27  & 0.43  & 0.30  &       &       &       &       &       &       &       &       &       &  \\
    Hungary & 0.26  & 0.46  & 0.28  &       &       &       &       &       &       &       &       &       &  \\
    Ireland & 0.26  & 0.48  & 0.27  &       &       &       &       &       &       &       &       &       &  \\
    Iceland & 0.25  & 0.46  & 0.29  &       &       &       &       &       &       &       &       &       &  \\
    Japan & 0.28  & 0.51  & 0.21  &       &       &       &       &       &       &       &       &       &  \\
    Kyrgyzstan & 0.25  & 0.47  & 0.27  &       &       &       &       &       &       &       &       &       &  \\
    Latvia & 0.25  & 0.49  & 0.26  &       &       &       &       &       &       &       &       &       &  \\
    Moldova & 0.24  & 0.50  & 0.26  &       &       &       &       &       &       &       &       &       &  \\
    Malta & 0.25  & 0.49  & 0.26  &       &       &       &       &       &       &       &       &       &  \\
    Netherlands & 0.24  & 0.64  & 0.12  &       &       &       &       &       &       &       &       &       &  \\
    Poland & 0.25  & 0.50  & 0.26  &       &       &       &       &       &       &       &       &       &  \\
    Portugal & 0.26  & 0.48  & 0.26  &       &       &       &       &       &       &       &       &       &  \\
    Serbia & 0.28  & 0.46  & 0.26  &       &       &       &       &       &       &       &       &       &  \\
    Slovakia & 0.19  & 0.59  & 0.21  &       &       &       &       &       &       &       &       &       &  \\
    Slovenia & 0.26  & 0.46  & 0.28  &       &       &       &       &       &       &       &       &       &  \\
    Sweden & 0.26  & 0.57  & 0.17  &       &       &       &       &       &       &       &       &       &  \\
    U.S.A. & 0.29  & 0.48  & 0.23  &       &       &       &       &       &       &       &       &       &  \\
    \midrule
    Mean  & 0.25  & 0.50  & 0.24  &       &       & 0.37  & 0.50  & 0.12  &       &       & 0.23  & 0.52  & 0.25 \\
    \bottomrule
    \end{tabular}%
    \begin{tablenotes}[para,flushleft]
    \vspace{3mm}
     \textit{Note:} This table reports the Shapley contribution of each component of Model \eqref{eq:final_estimation_model} to the predictions made by the model, averaged over the full period available for each country in proportions.
    \end{tablenotes}
    \end{threeparttable}
        }
  \label{tab:shapley}%
\end{table}%

To obtain a meaningful decomposition of the predictions made by Model \eqref{eq:final_estimation_model}, we follow the interpretable machine learning literature on Shapley decomposition \citep[see, for instance,][p. 218]{christoph2020interpretable}.
Instead of decomposing the $R^2$, we first rewrite the model predictions as
\begin{equation*}\label{eq:predictions}
    \widehat{S}_{it} = (\hat{\delta}_{\hat{g}_it} - \bar{\delta}_{\hat{g}_i}) \times CS_{it} + \bar{\delta}_{\hat{g}_i} \times CS_{it} + \hat{\mu}_{i} + \hat{\omega}_{\hat{g}_it},
\end{equation*}
where $\bar{\delta}_{\hat{g}_i} = T^{-1} \sum_{t=1}^T \hat{\delta}_{\hat{g}_it}$.
Then, we decompose this prediction for each country $i$ using Shapley contributions.
This representation allows us to consider the contribution of the capital share, $\bar{\delta}_{\hat{g}_i} \times CS_{it}$, and the effect of considering the time-varying nature of the relationship, $(\hat{\delta}_{\hat{g}_it} - \bar{\delta}_{\hat{g}_i}) \times CS_{it}$, separately.
Define the set of time-varying components of the model as $\mathcal{V} = \{ (\hat{\delta}_{\hat{g}_it} - \bar{\delta}_{\hat{g}_i}) \times CS_{it}, \bar{\delta}_{\hat{g}_i} \times CS_{it}, \hat{\omega}_{\hat{g}_it} \}$.
The Shapley contribution of each component $v \in \mathcal{V}$ is then calculated as
\begin{equation*}\label{eq:shapley}
\phi_{m,it} = \sum_{\mathcal{S} \subseteq \mathcal{V} \backslash \{v\}} [\hat{f}_{it}(\mathcal{S} \cup \{ v \}) - \hat{f}_{it}(\mathcal{S})]/3,
\end{equation*}
where $\hat{f}_{it}(\cdot)$ represents the prediction made using its arguments which are subsets of the explanatory variables plus $\hat{\mu}_{i}$ for country $i$ and year $t$.
The Shapley contribution of $v = (\hat{\delta}_{\hat{g}_it} - \bar{\delta}_{\hat{g}_i}) \times CS_{it}$ is then interpreted as the contribution of the transmission coefficient, $v = \bar{\delta}_{\hat{g}_i} \times CS_{it}$ is the effect of the capital share, and $v = \hat{\omega}_{\hat{g}_it}$ is the effect of the labor income inequality.

Table \ref{tab:shapley} reports the average Shapley contribution of each component of Model \eqref{eq:final_estimation_model} in terms of proportions, averaged for each country.
In the last row of the table, we also report the mean of each column by group.
The first observation is that in all three groups, the capital share is the most important factor in explaining the increase in the income inequality observed in the last four decades.
For Group 1 and Group 2, the average proportion explained by the capital share is 50\% and for Group 3, it is 52\%.
In terms of the proportions explained by country, Group 1 and Group 3 seems to be quite homogeneous.
The maximum proportion explained by the capital share is 64\% (Netherlands, Norway) in these groups and the minimum is 43\% (Crotia).
The explanatory power of the capital share remains quite homogeneous in Group 2 as well, whereas the other components show very different power for this group.
Labor income inequality in particular had an income inequality reducing effect for two countries in this group (Türkiye, Uruguay) and the transmission coefficient explains up to 75\% of the increase in Uruguay.
Overall, for Group 2 the transmission coefficient remains an important factor in the increase of the income inequality of the last four decades. For the sample of countries in hand, the capital share is found to be the most important driver.

\section{Conclusion}\label{sec:conclusion}

This paper was concerned with the estimation of the effect of the capital share on income inequality.
First, we outlined a theoretical framework with the help of a structural equation linking the capital share to income inequality. 
Second, we developed an estimation strategy for macrodata which provides a reliable alternative to a decomposition-based approach using microdata, as these often exhibit systematic measurement errors and inconsistencies.
By using the capital share from national accounts together with the top income shares from the World Income Database, we ensured the consistency between theory and measurement, an issue that has been particularly problematic in previous cross-country studies.
Third, and most importantly, we proposed an empirical methodology that explicitly accounts for the structural relationship between the capital share and income inequality, while also capturing both cross-country heterogeneity and time-varying nature in this relationship.
To the best our knowledge, this paper is the first cross-country study to incorporate time variance in a formal manner to measure this link.

Our findings showed that, on average over all 56 countries in the sample and the last four decades, one percentage point increase in capital share increases the income inequality by 0.17 percentage points.
However, the relationship is found to be depending on the group of countries under consideration and strongly time-dependent for a large group of countries. 
The main results on the time-varying nature of the relationship can be can be summarized as follows: \begin{enumerate*}[label=(\roman*)]
\item The transmission coefficient from capital share to personal income inequality is stable in a group of countries which is mainly made of advanced economies, and it increases in the group made of mostly emerging economies.
We also determined a third group for which the transmission coefficient decreases.
\item In the first group, mostly composed of advanced economies with low inequality and capital shares, we observe a simultaneous increase in the concentration of capital and labor incomes at the top of the distribution until 2008, keeping the transmission coefficient relatively stable throughout the period.
This is in line with the findings of two studies focusing on two advanced economies, namely \citet{aaberge2018, atkinson-lakner2021}, which show that from the late 1980s to the mid-2000s, the association between capital and labor incomes increased in Norway and the United States respectively.
\item In the second group, composed mainly of emerging economies with higher inequality and capital shares,  capital income inequality rises significantly along with  the transmission coefficient, while labor income inequality decreases, highlighting the dominance of capital as a driver of inequality in these countries.
\item In the other two groups, which together comprise developed, transition, and resource-driven economies with low to moderate inequality and capital shares, labor income inequality increases almost in the entire period, indicating that, in these countries, the top earners receive increasingly higher labor incomes.
\item  Finally, the Shapley decomposition of the model predictions reveals that the main driver of the changes in income inequality has been the changes in the capital share over the last four decades.
The percentage of the average increase in the income inequality due to the variations in the capital shares is found to be around 50\% for most of the countries in the sample.
\end{enumerate*}



\printbibliography
\end{document}